\newif\ifieee
\ieeefalse

\ifieee
\documentclass[compsoc, conference, a4paper, 10pt, times]{IEEEtran}
\else
	\documentclass{article}
	\usepackage[vmargin=1in,hmargin={1in,1.5in},headsep=0.3in,headheight=81pt]{geometry}
	\usepackage[utf8]{inputenc}
\fi

\usepackage{cite}
\usepackage{amsmath,amssymb,amsfonts}
\usepackage{algorithmic}
\usepackage{graphicx}
\usepackage{textcomp}
\usepackage{xcolor}
\usepackage{booktabs}
\usepackage{skull}
\usepackage{multirow}
\usepackage[hidelinks]{hyperref}

\usepackage{enumitem}
\usepackage{pgfplots}
\usepackage{algorithm}
\usepackage{hyperref}  
\PassOptionsToPackage{hyphens}{url}
\usepackage{tikz}
\usepackage{breakurl}
\usepackage{subcaption}

\usepgfplotslibrary{statistics}
\pgfplotsset{compat=1.11}

\newcommand{\NN}{\mathbb{N}}

\newcommand\xor{\mathbin{\char`\^}}
\renewcommand\xor{\mathbin{^\wedge}}

\newcommand\uxor{\xor}
\newcommand\unot{\mathord{\sim}}

\newcommand{\simpl}{\textit{GAMBA}}

\newcommand{\simba}{\textit{SiMBA}}
\newcommand{\mixed}{mixed}

\newtheorem{definition}{Definition}
\newtheorem{theorem}{Theorem}

\begin{document}

\title{Simplification of General Mixed Boolean-Arithmetic Expressions: GAMBA
\ifieee\else
\footnote{Accepted for presentation at the \href{https://worma.gitlab.io/2023/}{2nd Workshop on Robust Malware Analysis} (WoRMA'23), co-located with the \href{https://eurosp2023.ieee-security.org/index.html}{8th IEEE European Symposium on Security and Privacy}, Delft, The Netherlands, July 3 - 7, 2023.}
\fi
}

\ifieee
\else
	\author{Benjamin Reichenwallner \& Peter Meerwald-Stadler \\ Denuvo GmbH \\ Salzburg, Austria}
	\date{}
\fi

\maketitle

\begin{abstract}

Malware code often resorts to various self-protection techniques to complicate analysis.
One such technique is applying Mixed-Boolean Arithmetic (MBA) expressions as a way to create opaque predicates and diversify and obfuscate the data flow.

In this work we aim to provide tools for the simplification of nonlinear MBA expressions in a very practical context to compete in the arms race between the generation of hard, diverse MBAs and their analysis. The proposed algorithm \simpl{} employs algebraic rewriting at its core and extends \simba{}~\cite{simba}. It achieves efficient deobfuscation of MBA expressions from the most widely tested public datasets and simplifies expressions to their ground truths in most cases, surpassing peer tools.


\end{abstract}

\section{Introduction}

Mixed Boolean-arithmetic (MBA) expressions, which have been introduced in the year 2006 by Zhou et al.~\cite{zhou}, are a commonly used technique in code obfuscation. Their use in malware samples and various digital rights management (DRM) implementations is documented in the literature \cite{okane11,schrittwieser17,mougey14}; see \cite{mba-blast} for a detailed analysis of MBA usage in malware. They are believed to be one of the strongest-known code obfuscation techniques \cite{eqsat}.
In an effort to conceal secret information like data and algorithms, basic expressions like constants are transformed into mixed Boolean-arithmetic expressions that are semantically equivalent. This results in an artificial increase in code complexity to obfuscate the code and make it less comprehensible. It is typically assumed that the resultant complex expressions cannot be easily simplified back to their original form. 

However, recent research~\cite{simba} suggests that all linear MBAs can be solved in a straightforward way. This work extends this finding and contributes a practical algorithm for the simplification of nonlinear MBAs.

\subsection{Prior work}

Due to the fact that MBA expressions incorporate both logical and arithmetic operations that are not well compatible~\cite{sspam}, they cannot be genuinely resolved using SAT solvers or mathematical tools that are intended to handle solely logical or arithmetic expressions, resp. 

In recent years a significant number of tools specifically dedicated to their deobfuscation have been published~\cite{eqsat}. They use various techniques such as pattern matching (e.g., \mbox{SSPAM~\cite{sspam}}), neural networks (e.g., NeuReduce~\cite{neureduce}), bit-blasting (e.g., Arybo~\cite{arybo}), stochastic program synthesis (e.g., Stoke~\cite{stoke}, Syntia~\cite{syntia} and Xyntia~\cite{xyntia}), synthesis-based expression simplification (e.g., QSynth~\cite{qsynth} and msynth~\cite{msynth}), as well as a family of algebraic methods (e.g. \textit{MBA-Blast} \cite{mba-blast}, \textit{MBA-Solver} \cite{mba-solver}, \textit{MBA-Flatten} \cite{mba-flatten}, \simba{}~\cite{simba}).

The first step in code analysis often is symbolic execution: locating and extracting code from a malware sample and turning it into an abstract syntax tree (AST) \cite{qsynth,syntia,schrijver21}. In this work, we are not concerned with the many complications involved in lifting binary code, identifying an expression, etc. due to anti-debug/-trace techniques and program analysis limitations. We assume to have a pair of expressions $(e,e^\star)$, where $e$ is the complicated MBA expression and $e^\star$ the corresponding simple, semantically equivalent ground truth. The aim is to simplify $e$ in order to facilitate program analysis -- ideally back to $e^\star$. 
This setup is simpler compared to the scenario considered in program synthesis \cite{qsynth,syntia,msynth}, where obfuscated code (including control-flow problems) is taken as an input.

Zhou et al.~\cite{zhou} were also the first to describe how to generate random \textbf{linear} MBAs which are equivalent to a -- usually simple -- target expression $e^\star$ via a solution of a randomly generated equation system based on the finding that a bitwise expression is fully determined by its values in the set $B=\{0,1\}$ of zeros and ones. Another construction is an iterative application of rewriting rules from a given codebook, mapping simple expressions to more complex ones, see, e.g., \cite{loki}. To make MBAs more resistant to deobfuscation, additional encoding (e.g., using \textit{Tigress}~\cite{tigress}), invertible functions or point functions~\cite{zhou, loki} can be applied. These methods may create significant challenges for many simplification tools due to the potential introduction of large constants. Both generation approaches can in general also be used for generating \textbf{nonlinear} MBAs.
Obviously, a codebook may also be used for the simplification of MBAs. While MBA expressions may not be immediately present in the codebook, an SMT solver can be utilized to check for equivalency against the simpler MBAs listed. 

In 2021, Liu et al.~\cite{mba-blast, mba-solver} pointed out that Zhou et al.'s approach can be inverted too, pathing the way for surprisingly simple algebraic simplification of linear MBAs. This finding, which is actually already stated in Zhou et al.'s paper~\cite{zhou}, but remained unnoticed due to a mistake, is leveraged by the simplifiers \textit{MBA-Blast}~\cite{mba-blast}, \textit{MBA-Solver}~\cite{mba-solver}, \textit{MBA-Flatten}~\cite{mba-flatten} and \textit{SiMBA}~\cite{simba} operating on \textbf{linear} MBAs. These tools outperform other existing tools significantly for this class of MBAs when it comes to simplification success and runtime.

For the analysis of binary code, interactive disassemblers and decompilers are used. The \textit{SiMBA} algorithm has been integrated in Hex-Ray's \textit{gooMBA} plugin \cite{goomba} for \textit{IDA Pro}, to help turn complex MBA expressions given as IR code into simple terms. In a practical setting, fast, correct, easy-to-use and potent expression simplification is highly desirable. Early algorithms tend to be too slow (\textit{Arbyo}, \textit{SSPAM}, \textit{Syntia}), some require additional input (e.g., \textit{MBA-Solver} requires subexpressions), depend on training data or codebooks (\textit{NeuReduce}, \textit{SSPAM}), are nondeterministic (due to sampling of input data, e.g. \textit{QSynth}, \textit{Syntia}) or struggle with certain expressions (e.g., with nonlinear expressions or expressions with more than $5$ variables).

While linear MBAs have been shown to be easily solvable in general, it is still an open question whether more generic MBAs can constitute a solid obfuscation technique. Although the papers describing \textit{MBA-Solver}~\cite{mba-solver} and \textit{MBA-Flatten}~\cite{mba-flatten} claim to be capable of simplifying polynomial and even nonpolynomial MBAs, the published algorithms do not fully support this. They restrict the representation of input MBAs\footnote{E.g., the user has to know whether an MBA is linear, polynomial or nonpolynomial; the number of variables is limited to $4$; variables have to use prescribed variable names; a polynomial input MBA is required to be a sum of monomials; each term's first factor has to be a constant.}, and for a nonpolynomial input MBA it is required to provide a subexpression whose replacement by a variable makes the MBA a polynomial one, which is clearly impractical. Moreover, they do not output simplest solutions, but only ones using a set of bitwise base expressions.

\subsection{Contribution}

In this paper, we turn our attention to the simplification of \textbf{nonlinear} MBAs, using \simba{} as the core tool; we review preliminaries in Section~\ref{sec:prelim}. Although \simba{} is meant to operate on linear MBAs, it can in fact correctly solve all MBAs which are reducible to linear ones (see Section~\ref{sec:reducible}). 
We propose four improvements to \simba{} in Section~\ref{sec:ext} to make it more flexible.
Then we describe a more involved algorithm called \simpl{} (for \textit{Generalized Advanced MBA} simplifier) for nonlinear MBAs (Section~\ref{sec:deobf}) based on an iterative application of \simba{}, in combination with additional tricks and, critically, a substitution of nonlinear subexpressions by variables. 
While we cannot claim to be able to solve all MBAs with the current implementation, evaluation with publicly available datasets shows very promising results (Section~\ref{sec:results}). 



\section{Preliminaries}
\label{sec:prelim}

\subsection{Mixed Boolean-Arithmetic Expressions}\label{subsec:mbas}

Mixed Boolean-arithmetic expressions use logical (or bitwise, resp.) operators as well as arithmetic ones. 
While logical operators basically operate on $B=\{0,1\}$, i.e., single bits, bitwise operations are equivalently applied to all bits of $n$-bit words in $B^n$ for any fixed $n \in \NN$. This connection is of great importance, as it builds the foundation for the generation as well as the algebraic simplification of linear MBAs~\cite{zhou}.

As in~\cite{simba}, we prefer the notion of bitwise expressions which fits better to our context. That is, we use the operators $\&$ (bitwise conjunction), $\xor$ (bitwise exclusive disjunction), $\vert$ (bitwise inclusive disjunction) and $\sim$ (bitwise negation) rather than $\land$ (logical conjunction), $\oplus$ (logical exclusive disjunction), $\lor$ (logical inclusive disjunction) and $\lnot$ (logical negation).

The most popular classes of MBAs are that of linear and polynomial ones, resp. We first reiterate the definition of a polynomial MBA as in~\cite{simba}.

\begin{definition}\label{def:poly}
	Let $B = \{0,1\}$ and $n,t \in \NN$. A \textit{polynomial mixed Boolean-arithmetic expression} with values in $B^n$ and $t$ variables is a function $e: \left(B^n\right)^t \to B^n$ of the form $$e\left(x_1,\ldots,x_t\right) = \sum_{i\in I} a_i \prod_{j\in J_i} e_{ij}\left(x_1,\ldots,x_t\right),$$ where $I \subset \NN$ and $J_i\subset \NN$, for $i\in I$, are index sets, $a_i \in B^n$ are constants and $e_{ij}$ are bitwise expressions of variables $x_1,\ldots,x_t \in B^n$ for $j\in J_i$ and $i \in I$.
\end{definition}

As is easy to see and already noted in~\cite{simba}, a linear MBA is a special kind of a polynomial one:

\begin{definition}\label{def:linear}
	Let $B = \{0,1\}$ and $n,t \in \NN$. A \textit{linear mixed Boolean-arithmetic expression} with values in $B^n$ and $t$ variables is a function $e: \left(B^n\right)^t \to B^n$ of the form $$e\left(x_1,\ldots,x_t\right) = \sum_{i\in I} a_i e_i\left(x_1,\ldots,x_t\right),$$ where $I \subset \NN$ is an index set, $a_i \in B^n$ are constants and $e_i$ are bitwise expressions of $x_1,\ldots,x_t$ for $i \in I$.
\end{definition}

For instance, the MBA $x+(x\& y) - 2(x|y) + 42$ is linear, while the MBA $y(x\uxor y) - (x\& y)^2 - 1$ is polynomial, but not linear.

These notions were first introduced by Zhou et al.~\cite{zhou}. In this paper, we additionally use the term \textit{general MBA} for MBAs which are not necessarily polynomial. Here we restrict on those using the bitwise operations $\&$, $|$, $\uxor$ and $\unot$ as well as additions, subtractions, multiplications and powers (and implicitly left shifts, which can be written as multiplications of powers of $2$).

There are two possible reasons why such an MBA is nonpolynomial:

\begin{enumerate}
	\item It incorporates powers with nonconstant MBAs in their exponents, e.g., $3x^y + x + 17$. It seems obvious to us to call it an \textit{exponential} MBA in this case.
	\item It contains nontrivial constants or arithmetic operations in bitwise operations, e.g., $5+(x|3) - (5\& y)$. In this case we call it \textit{\mixed}.
\end{enumerate}

\begin{figure}[h]
	\centering
	\begin{tikzpicture}
		\draw[fill=gray!10] (1.15,0) ellipse (3.6cm and 1.8cm);
		\node[] at (2, -1.4) {\small general};
		
		\draw[fill=gray!10] (0,0) ellipse (2cm and 1cm);
		\node[] at (0, -0.7) {\small exponential};
		\draw[fill=gray!20] (0,.2) ellipse (1.4cm and 0.7cm);
		\node[] at (0, -0.25) {\small polynomial};
		\draw[fill=white] (0,.3) ellipse (0.8cm and 0.4cm);
		\node[] at (0, 0.3) {\small linear};
		
		\draw[fill=gray!10] (3.3,0) ellipse (1cm and 0.5cm);
		\node[] at (3.3, 0) {\small mixed};
		
		\draw (3.7, 1.6) -- (3.3, 0.3);
		\draw (3.5, 1.6) -- (3.2, 1.1);
		\draw (3.3, 1.6) -- (1.7, 0);
		\node[] at (3.7, 1.8) {\small nonpolynomial};
		
		\draw (1.45, 2.0) -- (2.8, 0.3);
		\draw (1.25, 2.0) -- (1.5, 1.4);
		\draw (1.05, 2.0) -- (1.0, 0.77);
		\draw (0.85, 2.0) -- (0.5, 0.7);
		\node[] at (1.15, 2.2) {\small nonlinear};
	\end{tikzpicture}
	
	\caption{MBA types}\label{fig:mba_types}
\end{figure}
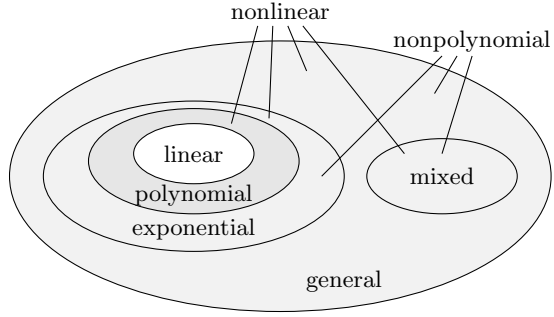

See Figure~\ref{fig:mba_types} for a visualization. By ``nontrivial'' constants, we mean constants that are neither $0$ nor $-1$ since those are the counterparts to the logical truth values. 

In their paper~\cite{zhou}, Zhou et al. describe a fundamental relation between Boolean and bitwise expressions that paves the way for a surprisingly simple, but popular method for generating linear MBAs:

\begin{theorem}\label{thm:zhou}
	Let $n, s, t \in \NN$, $x_i$ variables over $B^n$ for $i=1,\ldots,t$ and $e_j: \left(B^n\right)^t \to B^n$ bitwise expressions on these variables for $j=1, \ldots, s$. Let $$e\left(x_1,\ldots,x_t\right) = \sum_{j=1}^s a_j e_j\left(x_1,\ldots,x_t\right)$$ be a linear combination of these bitwise expressions with coefficients $a_j \in B^n$ for $j=1,\ldots,s$ and hence a linear MBA. Furthermore let, again for $j=1,\ldots,s$, $\overline{e}_j: B^t \to B$ be the logical expression corresponding to $e_j$. Enumerate the possible combinations of zeros and ones for the variables by $B_t = \{b_1, \ldots, b_{2^t}\}$ arbitrarily, but fixed, and let $A = \left(v_{ij}\right) \in B^{2^t \times s}$ be the matrix of truth values of the $\overline{e}_j$'s with $v_{ij} = \overline{e}_j(b_i)$.
	
	Then $e \equiv 0$ if and only if the vector $Y=Y_a = (a_1,\ldots,a_s)^T$ is a solution of the linear equation system $AY = o$ over $B^n$, where $o = (0,\ldots,0)^T$ is the zero vector in $B^{2^t}$.
\end{theorem}

According to this theorem, a linear MBA $e$ is equivalent to a bitwise expression $\tilde e$ on whole $B^n$, for any $n\in\NN$, if and only if they are equivalent on $B$. As will be noted in Theorem~\ref{thm:simba}, this also holds if $\tilde e$ is a linear MBA itself.

This builds the foundation for our simplifier \simba{} for linear MBAs and may also do so for generating general MBAs. To our knowledge, there is no obvious way how to define a similar approach for generating nonlinear MBAs via just an evaluation on inputs in $\{0,1\}$. An example showing that these values are not unique for nonlinear MBAs in general is easy to find. E.g., the polynomial MBA $x^2$ is equivalent to the linear MBA $x$ on $\{0,1\}$.

\subsection{Simplification of Linear Mixed Boolean-Arithmetic Expressions}

Recent research has shown that linear MBAs are for sure those which are easiest to solve. This is why most existing MBA simplifiers are especially successful with these MBAs or even only allow them as inputs. From the mentioned algebraic tools, \textit{MBA-Blast}, \textit{MBA-Solver} and \simba{} are all based -- more or less directly -- on Theorem~\ref{thm:zhou} while \textit{MBA-Flatten} uses a different approach to transform an input expression into a linear combination of conjunctions.

One main difference between \simba{} on the one hand and \textit{MBA-Blast} as well as \textit{MBA-Solver} on the other hand is that the latter use Theorem~\ref{thm:zhou} directly to transform each bitwise expression into a linear combination of bitwise expressions or a so-called \textit{signature vector} and then combine the results, \simba{} uses slightly more involved insights that allow to evaluate a linear MBA as a whole on $B=\{0,1\}$ and directly derive a first solution. We restate the underlying theorem~\cite{simba}:

\begin{theorem}\label{thm:simba}
	Let $e$ and $f$ be linear MBAs over words of the same length $n$ and let $t\in\NN$ be their (maximum) number of variables. Then $e \equiv f$ if and only if $e(b) = f(b)$ for each possible combination $b$ of in total $t$ zeros and ones.
\end{theorem}

What all these tools have in common is that they derive a linear combination of base bitwise expressions as a candidate for a solution. One possible base is the set of all conjunctions, e.g., $\{x, y, x\&y, -1\}$ for two variables, but there are multiple candidates and, e.g., \textit{MBA-Solver} uses a more complex basis. Therefore, the tools try to refine candidate solutions in order not to miss particularly simple solutions. While \textit{MBA-Blast} and \textit{MBA-Solver} check whether there is an equivalent expression using only one term, \simba{} guarantees to find the simplest solution for all inputs using two variables and for most inputs using three variables.

While the usage of \textit{MBA-Blast} and \textit{MBA-Solver} is restricted to MBAs with up to four variables due to their usage of lookup tables, with a number of entries which grows hyperexponentially following the formula $2^{2^t}$ for a number $t$ of variables, this insufficiency is eliminated by \textit{MBA-Flatten}, paying with a higher runtime and the lack of the possibility to find simplest solutions. In contrast, \simba{} is implemented fully generically and hence usable for an arbitrary number of variables, finding simple solutions for all inputs which are reducible to MBAs with at most three variables. We will explain in Section~\ref{sec:extension_simba} how this can be made even more flexible.

\section{Simplification of MBAs Reducible to Linear Ones Using \simba}
\label{sec:reducible}

Although \simba{} is mainly targeting linear MBAs, its usage is not restricted to those. Since it evaluates expressions directly on combinations of zeros and ones, its results are fully independent of the input expressions' representations. Unless \simba{} is instructed to neglect nonlinear inputs, it provides a result for any input MBA.

On the one hand, it is obvious that not every MBA is reducible to a linear one. But on the other hand, it is similarly easy to see that for each arbitrary MBA one can construct a linear one which corresponds to it for values in $\{0,1\}$. Consequently, and as already motivated in Subsection~\ref{subsec:mbas}, \simba's results are incorrect if and only if the input MBA has no equivalent representation as a linear MBA.

As a consequence, it is meaningful to run \simba{} on (not necessarily linear) MBAs whenever it is known that they are reducible to linear ones, i.e., their ground truths are linear. Since it can be assumed that a large portion of MBAs in practice are generated from linear ones, this extends \simba's field of application significantly.

Without any knowledge about the ground truth, one can still run the simplifier, but the result has to be verified. This can be done, e.g., via an evaluation of both expressions or their difference for values in $B^n$. However, this might take too long for larger $n$, and a check for just a subset of possible argument combinations will not give $100\%$ certainty.

Note that, e.g., a polynomial MBA with $t$ variables is in general not a polynomial in these variables. It is well a polynomial in bitwise expressions and can be transformed into a polynomial in at most $2^t$ base bitwise expressions, but when replacing the latter by variables, the information about their interdependence is lost. Hence, it is not obvious how basic analytic techniques can be leveraged to find roots of polynomial MBAs.

For example, consider the polynomial MBA \begin{align*}
	e_1 &= (-\unot (x|y) + (x|\unot y))(-(x\uxor y) - \unot (x\uxor y)) \\
	&+ (-2\unot (y|x) + \unot x + \unot (y\uxor x))(-\unot y - y)
\end{align*} and the exponential MBA \begin{align*}
	e_2 &= (-\unot (x|y) + (x|\unot y))^{-(x\uxor y) - \unot (x\uxor y)} \\
	&+ (-2\unot (y|x) + \unot x + \unot (y\uxor x))^{-\unot y - y},
\end{align*} which can both be simplified to $x+y$.\footnote{The second factor of each term in $e_1$ vanishes when writing negations $\unot X$ as $-X-1$, and equivalently for the exponents in $e_2$.} For another polynomial MBA \begin{align*}
	e_3 &= (x\&y)(x|y)+(x\&\unot y)(\unot x\&y),
\end{align*} \simba{} would output $\tilde e_3 = x\&y$ as a result. It is easy to verify that $\tilde e_3$ is equivalent to $e_3$ on $\{0,1\}$, but we have, e.g., that $e_3(1,2) = 2$, while $\tilde e_3(1,2) = 0$. In fact, it is not possible to transform $e_3$ into a linear MBA.

\section{Extending the Flexibilty of \simba}\label{sec:extension_simba}
\label{sec:ext}

We want to use \simba{} as a utility for the simplification of general MBAs (Section~\ref{sec:deobf}). For this purpose, we perform the following adaptations.

\subsection{Complexity Metrics}\label{subsec:metrics}

\simba{} finds for most expressions with two or three variables a solution with a minimal number of terms \cite{simba}. Obviously, such a solution is not necessarily the intuitively simplest solution. For instance, the expressions $e_4 = 2((y\&\unot z)|(x\&(y|\unot z)))-(x\uxor y\uxor z)$ and $\tilde e_4 = x+y-z$ are equivalent. While the former has fewer terms, the latter is intuitively the one which would be expected as a simplification result.

In general, the decision which solution is the ``simplest'' may depend on the type of application as well as the user's perspective. Hence, we extend \simba{} to choose a metric to guide decisions. Some possible metrics are based on the representation of an expression as an AST, which we prefer over a \textit{directed acyclic graph (DAG)}, in which equivalent nodes are merged. DAG nodes are impractical to us since coincident copies of a subexpression might be simplified in different ways.
We use the following metrics:

\begin{enumerate}
	\item \textbf{MBA alternation:} Proposed by Eyrolles~\cite{eyrolles} based on a DAG representation, it measures how often an expression alternates between bitwise and arithmetic operations. Thus, a purely bitwise or purely arithmetic expression has an MBA alternation of zero, independently of its number of terms.
	
	\item \textbf{Number of nodes in the AST representation:} Here we simply count the AST nodes, similarly to the DAG nodes, as suggested by Eyrolles~\cite{eyrolles}.
	
	\item \textbf{Number of terms:} This is the metric which we have used so far and which is easiest to compute and optimize for in our case.
	
	\item \textbf{String length:} The string length is easy to compute too, but does not give much insights into the structure of an MBA. Besides, it depends on the format of an MBA's string representation.
\end{enumerate}

If two considered MBAs have the same value for a metric, we apply a secondary metric to make a decision. Given a vector of results of an MBA for all possible combinations of arguments in $\{0,1\}$, it is easiest for us to determine the minimal number of terms we can achieve, while the other metrics' optimizers are harder to find. This extension implies that we have to compute more possible solutions and compare them.

In order to keep the number of inspected solutions small, we do not consider any solution with an equal or higher number of terms compared to the linear combination of conjunctions, i.e., the first candidate solution. We do not expect any such solution to minimize any of the metrics above, as the bitwise expressions in this linear combination are very simple.

Returning to the example above, the number of terms is the only metric which would prioritize $e_4$ over $\tilde e_4$; no other possible solution is additionally investigated.

Digging a bit deeper, the candidate solution $\tilde e_4$ is the initially determined linear combination of conjunctions, while $e_4$ is found in an attempt to decompose the input MBA's result vector $(0,1,1,2,-1,0,0,1)^\tau$ into a linear combination of truth value vectors of at most two bitwise expressions:
\begin{align*}
	2\,(0, 1, 1, 1, 0, 0, 0, 1)^\tau -(0, 1, 1, 0, 1, 0, 0, 1)^\tau.
\end{align*}


If we would consider all possible solutions consisting of three terms too, we would have to handle a surprisingly high number of different solutions: Even when fixing the bitwise expressions' coefficients to $1$, $1$ and $-1$, we can derive $3^5 = 243$ possible solutions, all using a different selection of three terms. In fact, even the set of feasible coefficients is only bounded by the number of bits, since, amongst other choices, $2, a$ and $-a-1$ would yield a positive number of solutions for any $a$.

\subsection{Additional Refinement Attempts}

\simba{}, as presented in~\cite{simba}, performs up to $8$ attempts to decompose a result vector in order to find a simple solution. We already indicated in the preceding subsection that we now perform in some occasions an exhaustive search for suitable bitwise expressions and coefficients that yield the same number of terms.

However, the algorithm still would not find solutions such as $x+\unot y$ or $\unot x + \unot y$. It is important to find such simple solutions in order to have an optimal chance to simplify complex general MBAs in the sequel. So we extend \simba{} by approaches to find linear combinations of a bitwise expression and a negated one, as well as linear combinations of two negated bitwise expressions.

Let $e$ be a linear MBA using $t \geq 2$ variables and $F = (e(b_1), \ldots,e(b_{2^t}))^\tau$ its vector of results when evaluated for all possible combinations $b_i$ of zeros and ones, where $b_i$ assigns the value $1$ to the variable $x_j$ if $i$'s remainder after a division by $2^j$ is larger than $2^{j-1}$. We extend \simba{} with the following two approaches to reduce the results' complexity:
\vspace{2mm}

\begin{enumerate}
 \item If $F$ has three or four distinct values, its first entry $a$ is nonzero and there are at most two values that are neither $a$ nor $2a$, we can express $e$ as a linear combination of an unnegated and a negated bitwise expression in the following cases:
 
	\begin{enumerate}
		\item If there is one such value $b$, we have two possible linear combinations with either coefficient $b-a$ or $b-2a$ for the unnegated bitwise expression and coefficient $-a$ for the negated one.
		\vspace{2mm}
		
	 \textit{Example for $t=2$:} The vector $(2, 2, 1, 4)^\tau$ can be decomposed into $$(0, 0, -1, 0)^\tau - 2\,(-1, -1, -1, -2)^\tau,$$ yielding the solution $-(\unot x\&y) -2\cdot \unot(x\& y)$, or $$(0, 0, -3, 0)^\tau - 2\,(-1, -1, -2, -2)^\tau,$$ yielding the solution $-3(\unot x\&y) -2\cdot \unot y$.
	 \vspace{2mm}
	 
  \item If there are two such values $b$, $c$ and, w.l.o.g., $c-b = a$, we have a linear combination with coefficient $b-a$ for the unnegated bitwise expression and coefficient $-a$ for the negated one.
		
		Note that in this case we might have multiple possible decompositions if $b=0$, since we can express $F$'s entries equal to $-a$ alternatively as $b-a$.
		\vspace{2mm}

		\textit{Example for $t=2$:} The vector $(-1,0,1,0)^\tau$ can be decomposed into $$2\,(0, 1, 1, 1)^\tau + \,(-1, -2, -1, -2)^\tau,$$ yielding the solution $2\,(x|y) + \unot x$.
	\end{enumerate}
	\vspace{2mm}
	
	\item If $F$ has three or four distinct values, its first entry $a$ is nonzero, there are exactly two values $b$, $c$ that are neither $a$ nor $2a$ and these sum up to $3a$, we can express $e$ as a linear combination of negated bitwise expressions with coefficients $a-b$ and $a-c$.
	
	The corresponding negated bitwise expressions' result vectors have value $-1$ where $F$ has either value $a$ or $c$ (or $b$, resp.) and value $-2$ where $F$ has either value $2a$ or $b$ (or $c$, resp.). The truth values of the bitwise expressions to be negated can then be derived by replacing $-1$ by $0$ and $-2$ by $1$.
	\vspace{2mm}
	
	\textit{Example for $t=2$:} The vector $(4,9,9,3)^\tau$ can be decomposed into $$(-1, -1, -1, -2)^\tau - 5\,(-1, -2, -2, -1)^\tau,$$ yielding the solution $\unot(x\&y) -5\cdot \unot(x\uxor y)$.
\end{enumerate}
\vspace{2mm}

The attentive reader might have noticed that by ignoring values equal to $2a$ we have neglected cases in which we can find solutions with just one term, as these are already found in a previous step of the algorithm -- see Subsection~3.2.4 of~\cite{simba}.

\subsection{Try to Split Expressions with More Variables}\label{subsec:more_vars}

If the initially determined linear combination of conjunctions uses more than three variables, this indicates that the input expression cannot be expressed using a fewer number of variables, and we thereby cannot use a lookup table for finding a simpler solution. Note that a lookup table for four variables would have $2^{2^4} = 65\,536$ entries, and one for five variables would already have $2^{2^5} = 4\,294\,967\,296$ entries.

In the next subsection we provide one possible workaround, but this increases the runtime for a higher number of variables. Yet we can avoid it in special cases: If we can nontrivially partition the initial solution's terms with respect to the occurring variables, i.e., such that the parts use disjunct sets of variables, we can handle these parts separately and combine the results. An optional constant may fit to either part, so we may have multiple possible solutions.

Consequently, for parts using at most three variables, we have the chance to run the usual procedure. For others we can apply the method described in the next subsection. As an example, consider the input expression
\begin{align*}
	&(a\&\unot b)+b-3((x\&\unot y)\uxor z)+3(\unot y|z)\\
	&\hspace{.5cm}-((x\&\unot y)\uxor \unot z)+4(\unot x|y) -4(\unot x\uxor (y\&z))\\
	&\hspace{.5cm}+(x\uxor (y\&\unot z))-x-2(\unot x\&(y|\unot z))\\
	&\hspace{.5cm}-2((x\&y)\uxor z),
\end{align*}
which can be transformed into the initial solution 
\begin{align*}
	&a+b-(a\&b)-2y-2z+2(x\&y)+2(x\&z)\\
	&\hspace{.5cm}+4(y\&z)-4(x\&y\&z).
\end{align*} 
This can be partitioned into three terms using the variables $a$ and $b$ and six terms using the variables $x$, $y$ and $z$. While the former can be simplified to $a|b$, the latter reduces to $-2(\unot x\&(y\uxor z))$. Hence, the result would be $$(a|b) -2(\unot x\&(y\uxor z)).$$

If a constant term exists in the initially determined linear combination, it fits to any partition. However, it is reasonable to choose the option which minimizes the used complexity metric.

\subsection{Creation of Base Bitwise Expressions}\label{subsec:bitwise_creation}

As mentioned, lookup tables of bitwise expressions grow fast with an increasing number of variables, hence we only use them for up to three variables. The peer tools \textit{MBA-Blast} and \textit{MBA-Solver} do so also for four variables, but as noted in~\cite{simba}, it slows the algorithms down drastically.

Fortunately, we can instantly create sufficiently simple bitwise expressions for any given vector of truth values using the \textit{Quine–McCluskey algorithm}~\cite{quine,mccluskey} to find a minimal disjunctive normal form, i.e., a disjunction of conjunctions of (possibly negated) variables that cannot be further reduced. Note that unfortunately its exponential runtime limits its usage too.

As an example, consider the truth value vector $(0,1,1,0,0,1,1,0)^\tau$ for four variables. The Quine-McCluskey algorithm would first create conjunctions for each $1$ in this vector, namely $x \& \unot y \& \unot z$, $x \& \unot y \& z$, $\unot x \& y \& \unot z$ and $\unot x \& y \& z$, and then merge the former two as well as the latter two  to finally get the disjunction $(x \& \unot y) | (\unot x \& y)$. Actually this can be further simplified to $x \uxor y$.

In order to identify chances to find simpler bitwise expressions as above, we perform refinements iteratively until nothing changes any more:
\begin{enumerate}
	\item \textit{Try to insert exclusive disjunctions:} For any bitwise subexpressions $X$ and $Y$, we can use the following patterns for a transformation into an exclusive disjunction $X\uxor Y$:	
	\begin{align*}
		(X\&\unot Y) | (\unot X \& Y) &\equiv X\uxor Y,\\
		(X|Y) \& (\unot X | \unot Y) &\equiv X\uxor Y.
	\end{align*}
	
	\item \textit{Potentially flip negations:} If a subexpression becomes simpler via flipping all its operands' negations, apply \textit{De Morgan's law}. Additionally, $\unot X \uxor \unot Y \equiv X \uxor Y.$
	\vspace{2mm}
	
	\item \textit{Try to factor out common subexpressions:} If a certain subexpression occurs in all operands of another subexpression, we apply the \textit{distributive law}. This includes the following well-known patterns:
	\begin{align*}
		(X\& Y) | (X\& Z) & \equiv X\& (Y | Z),\\
		(X| Y) \& (X| Z) & \equiv X| (Y \& Z).
	\end{align*}
\end{enumerate}

This list is not exhaustive and may be extended as desired. Note that other common rules such as the \textit{absorption laws} or the \textit{idempotence laws} have already implicitly been applied during the Quine-McCluskey algorithm.

\begin{algorithm}[h]
	\caption{Simplification of an MBA $e$ reducible to a linear one (extended \simba)}\label{alg:simba}
	\begin{enumerate}
		\item Determine linear combination $\tilde e \equiv e$ of conjunctions
		\item Identify the number $t$ of variables in $\tilde e$
		\item If $t \leq 3$:
		\begin{enumerate}
			\item\label{item:simba_refine} Try to find a simpler solution using table
		\end{enumerate}
		\item\label{item:simba_else} Else:
		\begin{enumerate}
			\item Determine partition $P$ of $\tilde e$ w.r.t. variables
			\item For all parts $p \in P$ with at most $3$ variables:
			\begin{enumerate}
				\item Try to find a simpler solution using table
			\end{enumerate}
			\item For all parts $p \in P$ with more than $3$ variables:
			\begin{enumerate}
				\item Try to find a simpler solution using bitwise creation as described in Subsection~\ref{subsec:bitwise_creation}
			\end{enumerate}
			\item Compose the results
		\end{enumerate}
	\end{enumerate}
\end{algorithm}

Algorithm~\ref{alg:simba} summarizes \simba's steps for simplifying MBAs that are reducible to linear ones. Here the whole branch~\ref{item:simba_else} is newly introduced, while step~\ref{item:simba_refine} has been extended.

\section{Simplification of General MBAs}
\label{sec:deobf}

In this section, we describe how we can use \simba{} in combination with additional steps to simplify MBAs that are not necessarily linear and not necessarily reducible to linear MBAs. As \simba, \simpl{} is written in \texttt{Python} without usage of packages such as \texttt{NumPy} or \texttt{SymPy} for nontrivial computations or simplification.

\subsection{Outline}

It is meaningful to operate on \textit{abstract syntax trees (ASTs)} rather than on strings. This helps identify variables, classify nodes as bitwise, linear or nonlinear subexpressions, compare expressions with more flexibility (e.g., not necessarily requiring a coincident order of operands), as well as extract, modify and reintegrate subexpressions.

Each node of an AST is either a variable, a constant or an operator with a certain number of operands. That is, each leaf node is a variable or a constant. The following operators, ordered by precedence, suffice for our purposes: power, bitwise negation ($\unot$), multiplication ($\cdot$), sum ($+$), conjunction ($\&$), exclusive disjunction ($\uxor$) and inclusive disjunction ($|$).

The strategy of \simpl{} is sketched in Algorithm~\ref{alg:general} which iteratively identifies linear subexpressions and simplifies them using \simba. In order to increase the chance of doing so, between these simplification runs, operations that support simplification as well as normalization are performed.

\begin{algorithm}[h]
	\caption{Simplification of a general MBA $e$ (\simpl)}\label{alg:general}
	\begin{enumerate}
		\item Parse $e$ into an AST $t$
		\item Repeat until convergence:
		\begin{enumerate}
			\item\label{item:refine} Refine $t$ as described in Subsection~\ref{subsec:node_operations}
			\item Identify linear subtrees of $t$ as described in Subsection~\ref{subsec:ident_linear}
			\item Try to factorize nonlinear sums as described in Subsection~\ref{subsec:factor}
			\item\label{item:simba} Apply \simba{} to linear subexpressions as sketched in Algorithm~\ref{alg:simba}
			\item Collect nodes for substitution
			\item For all combinations of those nodes:
			\begin{enumerate}
				\item Substitute these nodes by variables
				\item Apply steps~\ref{item:refine} to~\ref{item:simba}
				\item Back-replace these variables
				\item Refine $t$
			\end{enumerate}
		\end{enumerate}
		\item\label{item:polish} Polish $t$ for optimal representation
		\item Return a string representation of $t$
	\end{enumerate}
\end{algorithm}

In some cases it is nontrivial to retrieve linear parts of a nonlinear MBA if, e.g., products of linear subexpressions are expanded. To resolve that, \simpl{} incorporates a factorization procedure which tries to decompose sums of higher-order terms into factors.

Another main challenge for the simplification of MBAs that contain constants or arithmetic expressions within bitwise operations is to get rid of the latter or any other nonlinear subexpressions, if possible. This is done via a substitution logic, making subexpressions linear by substitution of nonlinear parts with temporary variables, simplifying and reinserting the substituted parts. 

The aim of step~\ref{item:polish} is to apply some standardization, including a deterministic order of operands in any kind of operation. This facilitates comparisons between simplified expressions.

This algorithm is a proof-of-concept and may have to be further adapted in order to be powerful enough to simplify (nearly) arbitrarily complex expressions.

\subsection{Node Operations for Refinement}\label{subsec:node_operations}

A refinement procedure is very crucial in order to establish invariants as well as to make sure that we can leverage a successful simplification of a subexpression. Imagine, e.g., that some expression cannot be further simplified because it has a subexpression like $(x | 3) \& \unot(x | 3)$, which is equivalent to $0$ and hence purely bitwise.

For normalization, we establish some trivial invariants such that all constant operands of a node are merged into one. 
%
%
%
Additionally we perform, amongst others, the following inspections in order to prepare the expression for simplification:

\vspace{2mm}
\noindent \textit{Applying logical rules in bitwise operations:} In conjunctions and exclusive as well as inclusive disjunctions, we get rid of all constants that are $0$ or $-1$, and resolve duplicate operands as well as operands that are inverse to each other. Furthermore \textit{De Morgan's law} may be applied to conjunctions or inclusive disjunctions.
\vspace{2mm}

\noindent \textit{Rearrangement of sums:} In sums, we collect terms that differ at most in constant factors and factor out common factors if possible.

\vspace{2mm}
\noindent \textit{Merging of powers:} In multiplications, powers are merged if they have the same base. This helps identify linear MBAs in exponents.\footnote{As an exponential MBA can be generated from a polynomial one by adding linear MBAs that are equivalent to $1$ as exponents and optionally splitting the arising powers, we try to restore such exponents.}

\vspace{2mm}
\noindent \textit{Eliminating or rewriting bitwise negations:} Nested negations are resolved, respecting the possibility that $\unot X$ is written as $-1-X$ or $-1(1+X)$ for any subexpression $X$. Furthermore, for any bitwise negation, either written explicitly using the negation operator or implicitly in arithmetic form, it is decided upon its context which representation increases the chance of simplification.

\vspace{2mm}
\noindent \textit{Flattening bitwise operations:} In some occasions a bitwise operation's complexity can be reduced via splitting it in terms, which is mainly meaningful in sums. This is, e.g., possible if an inclusive disjunction's operands are (or can be made) disjunct, including patterns such as $(X\&Y)|(X\uxor Y) \equiv (X\&Y) + (X\uxor Y)$ or even $X|(X\uxor Y) \equiv (X\&Y) + (X\uxor Y)$, which is equivalent to the former.\footnote{Since a bit of $X\uxor Y$ is $1$ if the corresponding bits in $X$ and $Y$ are different, it suffices to assume that they coincide in the inclusive disjunction's other operand, implying we can add a conjunction with $Y$.}

\vspace{2mm}
\noindent \textit{Factoring out from bitwise operations:} Powers of $2$ can be factored out from conjunctions, inclusive and exclusive disjunctions if they appear in all their operands. This can help get rid of constants in bitwise operations and hence make subexpressions linear. This corresponds to the pattern $2X\&2Y \equiv 2\,(X\&Y)$, and equivalent for the other operations. In fact we can even factor out a power of $2$ in cases where not all operands are divisible by that. Depending on the type of operation, we may have to compensate this by adding a remainder. Consider the following examples for subexpressions $X$ and $Y$ and a constant $a$ and remember that $\unot X \equiv -X-1$:
\begin{align*}
	\unot(2X)\&  2Y &\equiv 2\,(\unot X\&Y),\\
	\unot(2X)|  2Y &\equiv 2\,(\unot X|Y) + 1,\\
	(2a+1)\uxor  2X &\equiv 2\,(a\uxor X) + 1.
\end{align*}

\noindent \textit{Merging bitwise operations involving constants:} In sums, we may get rid of constants via merging bitwise operations with constants and, apart from that, coincident operators. This is particularly useful if the arising constants are $0$ or $-1$. The following rules hold for constants $a,b$ that have no $1$s in common in their binary representations:
\begin{align*}
	(a\& X) + (b\& X) &\equiv (a+b)\&X,\\
	(a| X) + (b| X) &\equiv ((a+b)|X) + X,\\
	(a\uxor X) + (b\uxor X) &\equiv 2\,(\unot(a+b)\&X) + a+b,\\
	(a| X) -(b\& X) &\equiv (\unot(a+b)\&X) + a,\\
	(a\uxor X) -2\,(b\& X) &\equiv 2\,(\unot(a+b)\&X) - X + a,\\
	(a\uxor X) + 2\,(b| X) &\equiv 2\,(\unot(a+b)\&X) + X + a + 2\,b.
\end{align*}

\noindent \textit{Merging bitwise operations involving inverse elements:} Similarly as explained above, terms of sums may be merged if the disjunct constants are replaced by inverse elements which are disjunct too. Some of the resulting patterns are well known:
\begin{align*}
	(X\& Y) + (\unot X\& Y) &\equiv Y,\\
	(X| Y) + (\unot X| Y) &\equiv -1 + Y,\\
	(X\uxor Y) + (\unot X\uxor Y) &\equiv -1,\\
	(X| Y) -(\unot X\& Y) &\equiv X,\\
	(X\uxor Y) -2\,(\unot X\& Y) &\equiv  X - Y,\\
	(X\uxor Y) + 2\,(\unot X| Y) &\equiv  \unot X+Y - 1.
\end{align*}

Note that we do not have to consider any optimization of linear expressions since \simba{} already outputs simplest expressions for them. However, the shown patterns also hold for non-bitwise subexpressions $X$ and $Y$.

\subsection{Identification of Linear Subexpressions}\label{subsec:ident_linear}

Before \simba{} can be applied to linear subexpressions, it is necessary to identify them. Additionally, purely bitwise expressions are identified as well. This is done in a straightforward way by induction:

\begin{itemize}[label=--]
	\item A variable is a bitwise expression.
	
	\item A constant node is considered bitwise if it is $0$ or $-1$ (corresponding to the logical constants), and linear otherwise.
	
	\item A subexpression corresponding to a bitwise operator (negation, conjunction, exclusive or inclusive disjunction) node is bitwise if all its operands are bitwise expressions, and nonlinear otherwise.
	
	\item A sum is nonlinear if it has a nonlinear term, and linear otherwise. In the former case, we can collect its linear terms which consequently form a linear subexpression.
	
	\item Being able to assume that at most one operand is constant, a product is nonlinear if it has more than two operands, at least one nonlinear operand or two operands and none of them is constant. Otherwise it is linear.
	
	\item Being able to assume that a power is never trivial, i.e., does not have the constant $1$ as exponent, it cannot be linear.
\end{itemize}

\subsection{Factorization}\label{subsec:factor}

The factorization of nonlinear sums is a crucial step for simplifying nonlinear MBAs whose linear parts are well-hidden by an expansion of their products. In such cases we cannot identify linear subexpressions easily, and the basic node operations as described in Subsection~\ref{subsec:node_operations} do not provide a solution. As an example, consider the MBA
\begin{align*}
	-x \cdot \unot(x|z) &- y \cdot \unot(x|z) - x\, (x\&\unot z)\\
	&- y \, (x\&\unot z) - xz - yz,
\end{align*}
which is in fact the product of the linear MBAs $x+y$ and $$-(\unot(x|z))-(x\& \unot z)-z,$$
whereas the latter is equivalent to a constant $1$.\footnote{This becomes evident after applying De Morgan's law to $\unot(x|z)$ to obtain $\unot x \& \unot z$, realizing that $(\unot x \& \unot z) + (x\& \unot z)$ is equivalent to $\unot z$ and writing the latter as $-z-1$.} That is, in order to simplify the MBA to $x+y$, we have to identify these factors.
\simpl{} iteratively finds simple factors which appear in a large number of terms of a sum, factors them out of them and splits the sum accordingly. In the sequel, these new terms can be collected if they only differ in the components that have been factored out.

In the previous example, $x$ and $y$ both appear in three terms. Hence, they would be factored out to obtain 
\begin{align*}
&x\, (-(\unot(x|z)) - (x\&\unot z) - z) \\
&\hspace{1cm} +y\, (-(\unot(x|z)) - (x\&\unot z) - z),
\end{align*}
these terms can be combined to 
\begin{align*}
	&(x+y)\, (-(\unot(x|z)) - (x\&\unot z) - z),
\end{align*}
and the latter factor vanishes by simplifying using \simba.

Before factorization, it may be necessary to expand products and powers and collect terms thereafter. Note that an MBA generator may, after generating linear MBAs for subexpressions, obscure those via expanding products and factorizing expressions into factors that cannot be simplified easily.

\subsection{Substitution of Subexpressions}

The techniques presented so far are usually not sufficient to simplify nonlinear MBAs, especially mixed ones, i.e., those containing constants or arithmetic operations within bitwise operations. In order to get rid of those, we apply some substitution logic. Our hope is to transform nonlinear subexpressions into linear ones via substitution of parts by variables, and that they get simpler via simplification using \simba{} and subsequent reinsertion of the substituted parts. As a simple example, consider the MBA $$((-x)\uxor y)-2\,((\unot -x)\&y)$$ which has no nontrivial linear subexpressions, but is in fact easily solvable by \simba{} after substituting $-x$ by a temporary variable, say, $X$. Then the expression $$(X\uxor y)-2((\unot X)\&y)$$ would resolve to $X-y$,\footnote{This can be seen after transforming $X\uxor y$ into $\unot((\unot X) \uxor y)$ and consequently into $-((\unot X) \uxor y)-1$, expanding $(\unot X) \uxor y$ to $((\unot X) | y) - ((\unot X) \& y)$, replacing  $(\unot X | y) + ((\unot X) \& y)$ by $\unot X + y$ and finally writing $\unot X$ as $-X-1$.} and after resubstitution to $-x-y$.

In some cases it might not be sufficient to only replace one subexpression because, e.g., subexpressions might remain nonlinear, but substituting a second subexpression might help. For instance, this is the case for the MBA \begin{align*}
	\unot x &+\unot (y-1)+2\\
	&+((-\unot x+1-1)|(-(\unot (y-1))-1)),
\end{align*}
which can only be simplified after a simultaneous substitution of $-(\unot x+1)-1$ by, say, $X$, and $-\unot(y-1)-1$ by, say, $Y$. 

In general, it is hard to decide on the right strategy since a simultaneous substitution may in some occasions hide too much of the interdependence between variables and subexpressions, so we collect all nodes which are meant to be substituted and run the substitution procedure on all possible subsets of this set. 

In fact, the example above shows that we have to identify subexpressions to be substituted even if they are not fully present: We actually substitute $\unot x + 1$ by $-X-1$ and $\unot (y-1)$ by $-Y-1$. This simplifies to $\unot(X\&Y)$ and to $-x|-y$ after back-substitution and refinement.

After substitution it might be necessary to do additional work regarding linear subexpressions which is usually done by \simba{}, but not in this case, when the interdependence between subexpressions is hidden by the substitution. In this course, e.g., we check whether terms of sums cancel out due to basic laws of logic.

\subsection{Remarks and Outlook}

We have described and implemented various techniques that support the simplification of MBAs of any kind, and the experiments below will suggest that this is a sophisticated starting point on the way to a powerful MBA simplifier. However, simplifying nonlinear MBAs is far from being straightforward and hence we can neither guarantee nor hope that this algorithm will be able to simplify all possible inputs. We consider it work in progress which can be improved with every input it fails to simplify. There will be further techniques to apply or transformation rules that can be implemented.

Apart from that, the current state of \simpl{} is not optimized regarding runtime. It may in some occasions consider subexpressions which are actually already optimally simplified. Moreover, we are aware of corner cases in which a tradeoff between performance and success probability has to be met, e.g., when a high number of possible solutions exist (see Subsection~\ref{subsec:metrics}) or the number of variables is large (see Subsection~\ref{subsec:more_vars}).

Besides, a high number of different nonlinear subexpressions, including constants, within larger subexpressions is challenging if they cannot be resolved applying transformation rules. In our substitution logic we have to substitute them by different variables, which increases the runtime. Our main goal is to show that it is well possible to simplify MBAs of any kind, and suggest one possible way how to achieve that.

\section{Experimental Results}
\label{sec:results}

All experiments are run on a Linux Mint 21 virtual machine on a single core of an Intel Core i7-12700K CPU at 3.6 GHz. The runtime was measured using Python 3.11 with the \texttt{time} package. Furthermore, we use $n = 64$ bits in all experiments unless otherwise noted.

We use six publicly available datasets. Table~\ref{tab:datasets} shows their numbers of expressions, expression types (linear, polynomial, nonpolynomial), numbers of variables (Vars), the average numbers of MBA alternations (Alt $\varnothing$), and the AST node count averages (Node $\varnothing$). All occurring nonpolynomial expressions are mixed, i.e., have constants or arithmetic operations within bitwise operations.

\begin{table}[h]
	\caption{Public MBA datasets\label{tab:datasets}}
	\begin{center}
		\begin{tabular}{crrrrr}
			\toprule
			Dataset & Expr. & Type & Vars & Alt $\varnothing$ & Node $\varnothing$ \tabularnewline
			\midrule
			NeuReduce \cite{neureduce} & $10\,000$ & linear & $2$ to $5$ & $7.9$ & $53.6$ \tabularnewline
			\midrule
			MBA-Obf. \cite{mba-obfuscator} & $1\,000$ & linear & $2$ to $3$ & $30.6$ & $267.7$ \tabularnewline
			linear & & & & &\tabularnewline
			\midrule
			MBA-Obf. \cite{mba-obfuscator} & $500$ & poly & $2$ & $ 18.7$ & $94.8$ \tabularnewline
			nonlinear & $500$ & nonpoly & $2$ & $26.1$ & $110.3$ \tabularnewline
			\midrule
			\multirow{3}{*}{Syntia \cite{mba-flatten}} & $182$ & linear & $1$ to $2$ & $1.6$ & $9.4$ \tabularnewline
			& $51$ & poly & $1$ to $3$ & $3.6$ & $17.0$ \tabularnewline
			& $267$ & nonpoly & $1$ to $3$ & $6.9$ & $27.6$  \tabularnewline
			\midrule
			\multirow{3}{*}{\scriptsize{MBA-Solver \cite{mba-solver}}} & $1\,000$ & linear & $1$ to $4$ & $9.1$ & $71.6$ \tabularnewline
			& $1\,000$ & poly & $1$ to $3$ & $9.5$ & $58.7$  \tabularnewline
			& $1\,000$ & nonpoly & $1$ to $3$ & $57.4$ & $306.1$ \tabularnewline
			\midrule
			QSynth EA \cite{qsynth}& $500$ & nonpoly & $1$ to $3$ & $77.6$ & $281.7$ \tabularnewline
			\bottomrule
		\end{tabular}
	\end{center}
\end{table}

Per dataset, we compute how many expressions can be simplified with \simba{} and \simpl{}, and compare with peer tools. Timeout is $60$~min unless noted otherwise and taken from references. Results reported are taken from the respective publications unless the tool name is written in \textit{italic} letters. Then they are self-generated. We do so for \textit{MBA-Flatten} since it is the closest and most recent related tool.

When presenting simplification results, $\equiv$ denotes that a complex MBA expression $e$ has been reduced to the corresponding ground truth $e^\star$. Moreover, $\approx$ indicates that the resulting expression is semantically equivalent to $e^\star$ (e.g. by simplifying their difference or proving equivalence using a SMT solver such as Z3~\cite{z3}). We use $\times$ when the result could not be proven to be correct, no result was given, or an error occurred.
The column \% indicates the percentage of successful experiments ($\equiv, \approx$) over the dataset.
Bold letters mark the best results per dataset.

Denote the simplification function by $S$. Given ($e$,$e^\star$) in the benchmarking scenario, a simplified expression $S(e)$ can be verified in several ways:
\renewcommand{\theenumi}{\roman{enumi}}
\begin{enumerate}
	\item It is identical to the ground truth, i.e., $S(e)\equiv e^\star$.
	\item It is identical to the simplified ground truth, i.e., $S(e)\equiv S(e^\star)$.
	\item\label{item:ver3} Their simplified difference $S(S(e) - S(e^\star))$ is zero, hence the result is semantically equivalent.
	\item It can be proven semantically equivalent using an SMT solver. For runtime performance reasons, the last check is often only computed for words $B^n$ with a limited number of bits, e.g., $n=4$ or $n=8$, allowing some incorrect results go undetected.
\end{enumerate}
\renewcommand{\theenumi}{\arabic{enumi}}

We put the focus on simplification success rather than runtimes. While we mention \simba's and \simpl's runtimes in order to be able to understand the complexity of, e.g., the additional refinement logic as well as iterative calls to \simba, we refer to~\cite{mba-blast} and ~\cite{simba} for runtime comparisons among nonalgebraic and algebraic tools, resp.

\subsection{NeuReduce}
The test dataset of \textit{NeuReduce}~\cite{neureduce} consists of $10\,000$ linear expressions with $2$ to $5$ variables. We observe that \simba{} and \simpl{} can simplify all expressions, even to the ground truth in all cases, see Table~\ref{tab:neureduce}. Note that \textit{MBA-Flatten} fails to handle expressions with $5$ variables.

\begin{table}[h]
\caption{NeuReduce dataset results \cite{neureduce}, timeout $40$ min\label{tab:neureduce}}
\begin{center}
\begin{tabular}{crrrrr}
	\toprule
Tool & $\equiv$ & $\approx$ & $\times$ & Timeout & \% \tabularnewline
\midrule
Arybo & $862$ & $0$ & $0$ & $9\,138$ & $8.6$\tabularnewline
SSPAM & $1\,420$ & $0$ & $0$ & $8\,580$ & $14.2$\tabularnewline
Syntia & $842$ & $734$ & $8\,424$ & $0$ & $15.8$\tabularnewline
NeuReduce & $7\,796$ & $28$ & $2176$ & $0$ & $78.2$\tabularnewline
\textit{MBA-Flatten} & $8\,560$ & $0$ & $1\,440$ & $0$ & $85.6$\tabularnewline
\textit{SiMBA} & $10\,000$ & $0$ & $0$ & $0$ & $\mathbf{100.0}$ \tabularnewline
\textit{GAMBA} & $10\,000$ & $0$ & $0$ & $0$ & $\mathbf{100.0}$ \tabularnewline
\bottomrule
\end{tabular}
\end{center}
\end{table}

The expressions of this dataset are simplified by \simba{} in $0.77$ ms (median: $0.74$ ms) and by \simpl{} in $8.30$ ms (median: $9.27$ ms) on the average. 

\subsection{MBA-Obfuscator}

The dataset consists of $1\,500$ linear, $1\,500$ polynomial and $1\,500$ nonpolynomial MBA expressions with $2$ to $4$ variables, but results are reported in~\cite{mba-obfuscator} only for the first $1\,000$ linear, and $1\,000$ nonlinear expressions (combining the first $500$ polynomial and nonpolynomial expressions).

\textit{MBA-Obfuscator}~\cite{mba-obfuscator} was the first proposal to generate diversified nonlinear MBA expressions. Due to the construction procedure, the nonlinear expressions share the same linear ground truth with the linear expressions. The linear expressions in the dataset are quite large: They comprise $267.7$ AST nodes on average, with a mean MBA alternation count of $30.6$ on average. 

\begin{table}[h]
	\caption{MBA-Obfuscator linear dataset results \cite{mba-obfuscator}\label{tab:mba-obf-linear}}
	\begin{center}
		\begin{tabular}{crrrrr}
			\toprule
			Tool & $\equiv$ & $\approx$ & $\times$ & Timeout & \% \tabularnewline
			\midrule
			Arybo & N/A & $569$ & $0$ & $431$ & $56.9$ \tabularnewline
			SSPAM & N/A & $386$ & $356$ & $258$ & $38.6$\tabularnewline
			Syntia & N/A & $97$ & $903$ & $0$ & $9.7$ \tabularnewline
			NeuReduce & N/A & $756$ & $244$ & $0$ & $75.6$ \tabularnewline
			MBA-Blast & N/A & $1\,000$ & $0$ & $0$ & $\mathbf{100.0}$ \tabularnewline
			\textit{MBA-Flatten} & $1\,000$ & $0$ & $0$ & $0$ & $\mathbf{100.0}$ \tabularnewline
			\textit{SiMBA} & $1\,000$ & $0$ & $0$ & $0$ & $\mathbf{100.0}$ \tabularnewline
			\textit{GAMBA} & $1\,000$ & $0$ & $0$ & $0$ & $\mathbf{100.0}$ \tabularnewline
			\bottomrule
		\end{tabular}
	\end{center}
\end{table}

The linear subset does not pose any problem to \textit{MBA-Blast}, \textit{MBA-Flatten}, \simba{} or \simpl{}, see Table~\ref{tab:mba-obf-linear}. The nonlinear subset is much harder: \textit{NeuReduce} can provide only one solution to an expression which is in fact still linear. \textit{MBA-Flatten} fails on several nonpolynomial MBAs in the dataset. Surprisingly, \simba{} succeeds in simplifying all expressions thanks to their linear ground truth, as explained in Section~\ref{sec:reducible}. \simpl{} in one case returns a slightly more complicated expression, but can still verify the result using verification strategy~\ref{item:ver3}.

\simba{} runs $2.47$ ms (median: $0.72$ ms) on the linear and $0.60$ ms (median: $0.62$ ms) on the nonlinear dataset on the average, while \simpl{} takes $18.41$ ms (median: $10.14$ ms) on the linear and $26.50$ ms (median: $15.12$ ms) on the nonlinear dataset on the average.

\begin{table}[h]
\caption{MBA-Obfuscator nonlinear dataset results \cite{mba-obfuscator}\label{tab:mba-obf-nonlinear}}
\begin{center}
\begin{tabular}{crrrrr}
	\toprule
Tool & $\equiv$ & $\approx$ & $\times$ & Timeout & \% \tabularnewline
\midrule
Arybo & N/A & $84$ & $0$ & $916$ & $8.4$\tabularnewline
SSPAM & N/A & $103$ & $192$ & $705$ & $10.3$\tabularnewline
Syntia & N/A & $98$ & $902$ & $0$ & $9.8$\tabularnewline
Neureduce & N/A & $1$ & $999$ & $0$ & $0.1$\tabularnewline
MBA-Blast & N/A & $147$ & $853$ & $0$ & $14.7$\tabularnewline
\textit{MBA-Flatten} & $953$ & $0$ & $47$ & $0$ & $95.3$\tabularnewline
\textit{SiMBA} & $1\,000$ & $0$ & $0$ & $0$ & $\mathbf{100.0}$\tabularnewline
\textit{GAMBA} & $999$ & $1$ & $0$ & $0$ & $\mathbf{100.0}$\tabularnewline
\bottomrule
\end{tabular}
\end{center}
\end{table}

\subsection{Syntia}

This dataset contains $500$ expressions with up to $3$ variables and was generated using the Tigress obfuscator~\cite{tigress}. Although several expressions are duplicates (only differ by variable name, $438$ expressions are unique), the dataset is used as a point of reference by many publications, e.g. \cite{qsynth}, \cite{mba-flatten}.

More than half of the expressions are nonpolynomial and $183$ of them are not reducible to a linear MBA. Consequently \simba{} cannot simplify the latter, see Table~\ref{tab:syntia}. While \textit{QSynth}, \textit{MBA-Flatten} and \simpl{} can simplify the entire dataset, only \simpl{}'s results are always identical to the ground truth.
Note that for solving the \textit{Syntia} dataset, \textit{MBA-Flatten} uses a non-generic, customized implementation based on knowledge about these MBAs' structure.

\begin{table}[h]
\caption{Syntia dataset results \cite{mba-flatten}\label{tab:syntia}}
\begin{center}
\begin{tabular}{crrrrr}
	\toprule
Tool & $\equiv$ & $\approx$ & $\times$ & Timeout & \%\tabularnewline
\midrule
SSPAM & N/A & $332$ & $168$ & $0$ & $66.4$\tabularnewline
Syntia & N/A & $369$ & $131$ & $0$ & $73.8$\tabularnewline
QSynth & N/A & $500$ & $0$ & $0$ & $\mathbf{100.0}$\tabularnewline
MBA-Blast & N/A & $416$ & $0$ & $84$ & $83.2$\tabularnewline
MBA-Solver & N/A & $454$ & $0$ & $46$ & $90.8$\tabularnewline
\textit{MBA-Flatten} & $302$ & $198$ & $0$ & $0$ & $\mathbf{100.0}$\tabularnewline
\textit{SiMBA} & $317$ & $0$ & $183$ & $0$ & $63.4$\tabularnewline
\textit{GAMBA} & $500$ & $0$ & $0$ & $0$ & $\mathbf{100.0}$\tabularnewline
\bottomrule
\end{tabular}
\end{center}
\end{table}

\simba{}'s average runtime for this dataset is $0.18$ ms (median: $0.10$~ms) and that of \simpl{} is $8.89$ ms (median: $7.67$~ms).

\subsection{MBA-Solver}

The dataset contains $1\,000$ linear, $1\,000$ polynomial and $1\,000$ nonpolynomial MBA expressions with up to $4$ variables. The nonpolynomial expressions are comparatively large: $57.4$ MBA alternations and $306.1$ AST nodes on average.

\begin{table}[h]
	\caption{MBA-Solver dataset results \cite{mba-flatten}\label{tab:mba-solver}}
	\begin{center}
		\begin{tabular}{crrrrrr}
			\toprule
			Tool & $\equiv$ & $\approx$ & $\times$ & Timeout & \%\tabularnewline
			\midrule
			SSPAM & N/A & $705$ & $320$ & $1\,975$ & $34.2$\tabularnewline
			Syntia & N/A & $437$ & $2\,563$ & $0$ & $14.6$\tabularnewline
			MBA-Blast & N/A & $1\,763$ & $0$ & $1\,237$ & $58.8$\tabularnewline
			MBA-Solver & N/A & $2\,899$ & $0$ & $101$ & $96.6$\tabularnewline
			\textit{MBA-Flatten} & $2\,500$ & $443$ & $0$ & $57$ & $98.1$\tabularnewline
			\textit{SiMBA} & $1\,757$ & $87$ & $1\,156$ & $0$ & $61.5$ \tabularnewline
			\textit{GAMBA} & $2\,998$ & $2$ & $0$ & $0$ & $\mathbf{100.0}$\tabularnewline
			\bottomrule
		\end{tabular}
	\end{center}
\end{table}

Simplification results are shown in Table~\ref{tab:mba-solver}. \textit{MBA-Solver} manages to solve $2\,899$ expression, but depends on subexpressions as additional input to steer the simplification process. \textit{MBA-Flatten} produces {2\,500} results equivalent to the ground truth and can additionally provide $443$ semantically equivalent solutions.
\simba{} can return results not only for linear expressions, but also for some nonpolynomial MBAs; yet it cannot solve any of the polynomial MBA expressions, as they are not reducible to linear ones.
\simpl{} can simplify all expression in the dataset, to an expression equivalent to the ground truth in almost all cases.

While \simba{}'s average runtime for this dataset is $10.39$ ms (median: $1.65$~ms), that of \simpl{} is $19.12$ ms (median: $11.96$~ms).

\subsection{QSynth EA}

This dataset contains $500$ nonpolynomial MBA expressions with up to $3$ variables. The expressions were generated with the \textit{Tigress} obfuscator \cite{tigress} by applying its \textit{EncodeArithmetic} (EA) transform. The expressions are large: $281.7$ AST nodes on average. Furthermore, this dataset's average number of  MBA alternations is $77.6$, the highest in this comparison.

Table~\ref{tab:qsynth-ea} shows that \simpl{} can simplify $98.4\%$ of these complex MBAs.
For $61$ expressions, the equivalence to the corresponding ground truths can be verified. It is worth to note that for the $8$ runs where the verification was not successful (using verification methods iii or iv), in fact, the expressions were drastically shorter as well. \simba{} can only simplify $45$ expressions, as the remaining ones have no linear ground truth. \textit{MBA-Flatten} is not included here since it cannot handle the occurring variable names, shift operators and the MBAs' general structure.

\begin{table}[h]
	\caption{QSynth EA dataset results \cite{qsynth}, timeout $1$ min\label{tab:qsynth-ea}}
	\begin{center}
		\begin{tabular}{crrrrrr}
			\toprule
			Tool & $\equiv$ & $\approx$ & $\times$ & Timeout & \% \tabularnewline
			\midrule
			QSynth & N/A & $354$ & $146$ & $0$ & $69.0$\tabularnewline
			\textit{SiMBA} & $45$ & $0$ & $455$ & $0$ & $9.0$ \tabularnewline
			\textit{GAMBA} & $431$ & $61$ & $8$ & $0$ & $\mathbf{98.4}$\tabularnewline
			\bottomrule
		\end{tabular}
	\end{center}
\end{table}

It takes \simba{} $0.81$~ms on the average (median: $0.59$~ms) to solve those expressions that it can solve. \simpl{} runs $134.18$ ms on the average (median: $53.95$~ms). Figure~\ref{fig:box} demonstrates the high variance of \simpl's runtimes when applied to the QSynth EA dataset. The maximum runtime is about $3\,082.2$~ms while most runs take fractions of a second. It is also notable that all equivalences in column $\approx$ were proved by \simpl{} itself, i.e., usage of Z3 was not required.

\pgfplotsset{every tick label/.append style={font=\scriptsize}}

\begin{figure}[h]
	\caption{Boxplot of \simpl's runtimes on QSynth EA}\label{fig:box}
\centering
\begin{tikzpicture}
	\begin{axis}
		[
		ytick=\empty,
		yticklabels=\empty,
		y post scale=0.2,major tick length=2.5,xlabel={\footnotesize$s$}, x label style={at={(axis description cs:1.0,-0.03)},anchor=north}
		]
		\addplot+[draw=black,nodes=black,fill=white,solid,mark=*,every mark/.append style={fill=white,},
		boxplot prepared={
			median=0.05396973,
			upper quartile=0.13490246,
			lower quartile=0.02076974,
			upper whisker=0.30610154,
			lower whisker=0.00762097
		},
		] coordinates{
			(0,0.3233195149950916)
			(0,1.0604229620075785)
			(0,0.5638370999949984)
			(0,0.43319529700966086)
			(0,0.4654790659988066)
			(0,0.5522970390011324)
			(0,0.417421257996466)
			(0,0.508209159001126)
			(0,0.3784615579934325)
			(0,1.172992902007536)
			(0,1.9782968539948342)
			(0,0.34445760199741926)
			(0,3.082182414000272)
			(0,0.5087316429999191)
			(0,0.7768351919949055)
			(0,0.3121200070017949)
			(0,0.4950694989965996)
			(0,0.6745436290075304)
			(0,0.39438568899640813)
			(0,0.37182064600347076)
			(0,1.0344195869984105)
			(0,0.3826259010093054)
			(0,0.43837587999587413)
			(0,0.43340407000505365)
			(0,0.328573184000561)
			(0,0.3891779650002718)
			(0,0.536467180994805)
			(0,0.8126738150021993)
			(0,0.4827861879894044)
			(0,0.7885446089931065)
			(0,0.33164804500120226)
			(0,0.43100732800667174)
			(0,0.3105494670016924)
			(0,0.6978533110086573)
			(0,0.33964772200852167)
			(0,0.36104479899222497)
			(0,0.43384905499988236)
			(0,0.4364195020025363)
			(0,0.4775508489983622)
			(0,0.9471020429919008)
			(0,0.5474974419921637)
			(0,0.3242334280075738)
			(0,1.7177113700017799)
			(0,0.5938483249919955)
			(0,0.392964093000046)
			(0,1.3188790639978833)
			(0,0.3664782349951565)
			(0,0.529998997997609)
			(0,0.40158546199381817)
			(0,0.3817245029931655)
			(0,0.34732255700509995)
			(0,0.7454941130126826)
			(0,0.3943895470001735)
			(0,0.8227568889997201)
			(0,0.5044895200117026)
			(0,0.3199376900010975)
			(0,0.321310800005449)
			(0,0.984635716988123)
			(0,0.6980529489956098)
		};
	\end{axis}
\end{tikzpicture}
\end{figure}
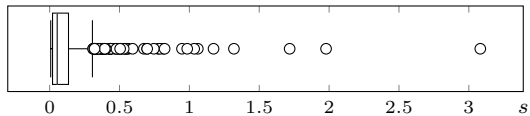

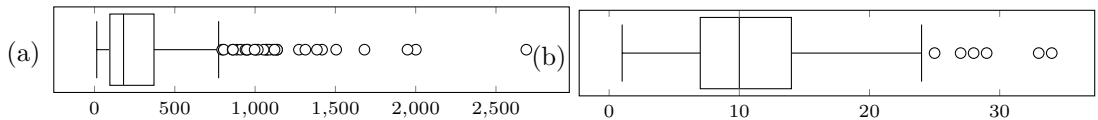
\begin{figure}[h]
	\caption{Boxplots describing the QSynth EA dataset: number of nodes of (a) $e$ and (b) ground truth $e^\star$}\label{fig:boxes}
\centering
\begin{subfigure}[b]{0.45\textwidth}
	\centering	
\begin{tikzpicture}
	\begin{axis}
		[
		ytick=\empty,
		yticklabels=\empty,
		y post scale=0.2,major tick length=2.5
		]
		\addplot+[draw=black,nodes=black,fill=white,solid,mark=*,every mark/.append style={fill=white,},
		boxplot prepared={
			median=181,
			upper quartile=370.5,
			lower quartile=95.75,
			upper whisker=773,
			lower whisker=14
		},
		] coordinates{
			(0,1417)
			(0,1113)
			(0,1681)
			(0,1073)
			(0,955)
			(0,795)
			(0,805)
			(0,1385)
			(0,1069)
			(0,2001)
			(0,1089)
			(0,885)
			(0,1273)
			(0,1137)
			(0,909)
			(0,860)
			(0,1949)
			(0,1003)
			(0,1137)
			(0,1121)
			(0,940)
			(0,862)
			(0,1053)
			(0,2689)
			(0,949)
			(0,1505)
			(0,1313)
			(0,1017)
			(0,997)
		};
	\end{axis}

	\node[] at (-0.4, 0.5) {(a)};
\end{tikzpicture}
\end{subfigure}
\begin{subfigure}[b]{0.45\textwidth}
\centering
\begin{tikzpicture}
	\begin{axis}
		[
		ytick=\empty,
		yticklabels=\empty,
		y post scale=0.2,major tick length=2.5
		]
		\addplot+[draw=black,nodes=black,fill=white,solid,mark=*,every mark/.append style={fill=white,},
		boxplot prepared={
			median=10,
			upper quartile=14,
			lower quartile=7,
			upper whisker=24,
			lower whisker=1
		},
		] coordinates{
			(0,29)
			(0,33)
			(0,34)
			(0,27)
			(0,28)
			(0,25)
		};
	\end{axis}

	\node[] at (-0.4, 0.5) {(b)};
\end{tikzpicture}
\end{subfigure}
%
\end{figure}

Since this is the most complex dataset, we additionally consider the complexities of the original MBAs as well of the results. Figure~\ref{fig:boxes} shows statistics on the numbers of nodes of the original expressions as well as of their ground truths, indicating that we have a variety of very differently complex expressions.

\begin{figure}[h]
\caption{Deviances in the numbers of nodes from ground truths of \textcolor{blue}{equivalent ($\approx$)} and \textcolor{red}{failed ($\times$)} runs of \simpl{} on the QSynth EA dataset}\label{fig:deviances}
\centering
	\centering
\begin{tikzpicture}
	\def\pairsSim{(12,25)(15,35)(12,13)(10,25)(13,39)(14,17)(20,22)(19,31)(23,94)(13,20)(13,30)(15,30)(15,46)(12,50)(20,39)(17,70)(14,30)(18,18)(12,26)(14,35)(15,27)(7,17)(15,20)(14,30)(13,18)(9,10)(11,22)(12,27)(17,18)(15,15)(16,53)(10,4)(13,32)(11,16)(14,17)(10,71)(14,61)(19,67)(14,16)(14,12)(15,29)(17,36)(20,27)(18,151)(13,72)(16,181)(27,69)(9,10)(13,29)(11,18)(13,33)(14,40)(21,29)(14,23)(15,22)(12,27)(15,31)(21,57)(19,40)(28,29)(6,22)}

	\pgfmathsetmacro{\t}{0.05};
	\pgfmathsetmacro{\r}{0.2};
	
	\begin{axis}[
		ytick=\empty,y post scale=0.5,
		yticklabels = \empty,]
		
		\addplot[blue, mark=none] coordinates {(28, 0*\t) (29, 0*\t)};
		\addplot[blue, mark=none] coordinates {(27, 1*\t) (69, 1*\t)};
		\addplot[blue, mark=none] coordinates {(23, 2*\t) (94, 2*\t)};
		\addplot[blue, mark=none] coordinates {(21, 3*\t) (29, 3*\t)};
		\addplot[blue, mark=none] coordinates {(21, 4*\t) (57, 4*\t)};
		\addplot[blue, mark=none] coordinates {(20, 5*\t) (22, 5*\t)};
		\addplot[blue, mark=none] coordinates {(20, 6*\t) (39, 6*\t)};
		\addplot[blue, mark=none] coordinates {(20, 7*\t) (27, 7*\t)};
		\addplot[blue, mark=none] coordinates {(19, 8*\t) (31, 8*\t)};
		\addplot[blue, mark=none] coordinates {(19, 9*\t) (67, 9*\t)};
		\addplot[blue, mark=none] coordinates {(19, 10*\t) (40, 10*\t)};
		\addplot[blue, mark=none] coordinates {(18, 11*\t) (18, 11*\t)};
		\addplot[blue, mark=none] coordinates {(18, 12*\t) (151, 12*\t)};
		\addplot[blue, mark=none] coordinates {(17, 13*\t) (70, 13*\t)};
		\addplot[blue, mark=none] coordinates {(17, 14*\t) (18, 14*\t)};
		\addplot[blue, mark=none] coordinates {(17, 15*\t) (36, 15*\t)};
		\addplot[blue, mark=none] coordinates {(16, 16*\t) (53, 16*\t)};
		\addplot[blue, mark=none] coordinates {(16, 17*\t) (181, 17*\t)};
		\addplot[blue, mark=none] coordinates {(15, 18*\t) (35, 18*\t)};
		\addplot[blue, mark=none] coordinates {(15, 19*\t) (30, 19*\t)};
		\addplot[blue, mark=none] coordinates {(15, 20*\t) (46, 20*\t)};
		\addplot[blue, mark=none] coordinates {(15, 21*\t) (27, 21*\t)};
		\addplot[blue, mark=none] coordinates {(15, 22*\t) (20, 22*\t)};
		\addplot[blue, mark=none] coordinates {(15, 23*\t) (15, 23*\t)};
		\addplot[blue, mark=none] coordinates {(15, 24*\t) (29, 24*\t)};
		\addplot[blue, mark=none] coordinates {(15, 25*\t) (22, 25*\t)};
		\addplot[blue, mark=none] coordinates {(15, 26*\t) (31, 26*\t)};
		\addplot[blue, mark=none] coordinates {(14, 27*\t) (17, 27*\t)};
		\addplot[blue, mark=none] coordinates {(14, 28*\t) (30, 28*\t)};
		\addplot[blue, mark=none] coordinates {(14, 29*\t) (35, 29*\t)};
		\addplot[blue, mark=none] coordinates {(14, 30*\t) (30, 30*\t)};
		\addplot[blue, mark=none] coordinates {(14, 31*\t) (17, 31*\t)};
		\addplot[blue, mark=none] coordinates {(14, 32*\t) (61, 32*\t)};
		\addplot[blue, mark=none] coordinates {(14, 33*\t) (16, 33*\t)};
		\addplot[blue, mark=none] coordinates {(14, 34*\t) (12, 34*\t)};
		\addplot[blue, mark=none] coordinates {(14, 35*\t) (40, 35*\t)};
		\addplot[blue, mark=none] coordinates {(14, 36*\t) (23, 36*\t)};
		\addplot[blue, mark=none] coordinates {(13, 37*\t) (39, 37*\t)};
		\addplot[blue, mark=none] coordinates {(13, 38*\t) (20, 38*\t)};
		\addplot[blue, mark=none] coordinates {(13, 39*\t) (30, 39*\t)};
		\addplot[blue, mark=none] coordinates {(13, 40*\t) (18, 40*\t)};
		\addplot[blue, mark=none] coordinates {(13, 41*\t) (32, 41*\t)};
		\addplot[blue, mark=none] coordinates {(13, 42*\t) (72, 42*\t)};
		\addplot[blue, mark=none] coordinates {(13, 43*\t) (29, 43*\t)};
		\addplot[blue, mark=none] coordinates {(13, 44*\t) (33, 44*\t)};
		\addplot[blue, mark=none] coordinates {(12, 45*\t) (25, 45*\t)};
		\addplot[blue, mark=none] coordinates {(12, 46*\t) (13, 46*\t)};
		\addplot[blue, mark=none] coordinates {(12, 47*\t) (50, 47*\t)};
		\addplot[blue, mark=none] coordinates {(12, 48*\t) (26, 48*\t)};
		\addplot[blue, mark=none] coordinates {(12, 49*\t) (27, 49*\t)};
		\addplot[blue, mark=none] coordinates {(12, 50*\t) (27, 50*\t)};
		\addplot[blue, mark=none] coordinates {(11, 51*\t) (22, 51*\t)};
		\addplot[blue, mark=none] coordinates {(11, 52*\t) (16, 52*\t)};
		\addplot[blue, mark=none] coordinates {(11, 53*\t) (18, 53*\t)};
		\addplot[blue, mark=none] coordinates {(10, 54*\t) (25, 54*\t)};
		\addplot[blue, mark=none] coordinates {(10, 55*\t) (4, 55*\t)};
		\addplot[blue, mark=none] coordinates {(10, 56*\t) (71, 56*\t)};
		\addplot[blue, mark=none] coordinates {(9, 57*\t) (10, 57*\t)};
		\addplot[blue, mark=none] coordinates {(9, 58*\t) (10, 58*\t)};
		\addplot[blue, mark=none] coordinates {(7, 59*\t) (17, 59*\t)};
		\addplot[blue, mark=none] coordinates {(6, 60*\t) (22, 60*\t)};
		\node at (28, 0*\t) [circle, scale=\r, draw=black,fill=black] {};
		\node at (29, 0*\t) [circle, scale=\r, draw=blue,fill=blue] {};
		\node at (27, 1*\t) [circle, scale=\r, draw=black,fill=black] {};
		\node at (69, 1*\t) [circle, scale=\r, draw=blue,fill=blue] {};
		\node at (23, 2*\t) [circle, scale=\r, draw=black,fill=black] {};
		\node at (94, 2*\t) [circle, scale=\r, draw=blue,fill=blue] {};
		\node at (21, 3*\t) [circle, scale=\r, draw=black,fill=black] {};
		\node at (29, 3*\t) [circle, scale=\r, draw=blue,fill=blue] {};
		\node at (21, 4*\t) [circle, scale=\r, draw=black,fill=black] {};
		\node at (57, 4*\t) [circle, scale=\r, draw=blue,fill=blue] {};
		\node at (20, 5*\t) [circle, scale=\r, draw=black,fill=black] {};
		\node at (22, 5*\t) [circle, scale=\r, draw=blue,fill=blue] {};
		\node at (20, 6*\t) [circle, scale=\r, draw=black,fill=black] {};
		\node at (39, 6*\t) [circle, scale=\r, draw=blue,fill=blue] {};
		\node at (20, 7*\t) [circle, scale=\r, draw=black,fill=black] {};
		\node at (27, 7*\t) [circle, scale=\r, draw=blue,fill=blue] {};
		\node at (19, 8*\t) [circle, scale=\r, draw=black,fill=black] {};
		\node at (31, 8*\t) [circle, scale=\r, draw=blue,fill=blue] {};
		\node at (19, 9*\t) [circle, scale=\r, draw=black,fill=black] {};
		\node at (67, 9*\t) [circle, scale=\r, draw=blue,fill=blue] {};
		\node at (19, 10*\t) [circle, scale=\r, draw=black,fill=black] {};
		\node at (40, 10*\t) [circle, scale=\r, draw=blue,fill=blue] {};
		\node at (18, 11*\t) [circle, scale=\r, draw=black,fill=black] {};
		\node at (18, 11*\t) [circle, scale=\r, draw=blue,fill=blue] {};
		\node at (18, 12*\t) [circle, scale=\r, draw=black,fill=black] {};
		\node at (151, 12*\t) [circle, scale=\r, draw=blue,fill=blue] {};
		\node at (17, 13*\t) [circle, scale=\r, draw=black,fill=black] {};
		\node at (70, 13*\t) [circle, scale=\r, draw=blue,fill=blue] {};
		\node at (17, 14*\t) [circle, scale=\r, draw=black,fill=black] {};
		\node at (18, 14*\t) [circle, scale=\r, draw=blue,fill=blue] {};
		\node at (17, 15*\t) [circle, scale=\r, draw=black,fill=black] {};
		\node at (36, 15*\t) [circle, scale=\r, draw=blue,fill=blue] {};
		\node at (16, 16*\t) [circle, scale=\r, draw=black,fill=black] {};
		\node at (53, 16*\t) [circle, scale=\r, draw=blue,fill=blue] {};
		\node at (16, 17*\t) [circle, scale=\r, draw=black,fill=black] {};
		\node at (181, 17*\t) [circle, scale=\r, draw=blue,fill=blue] {};
		\node at (15, 18*\t) [circle, scale=\r, draw=black,fill=black] {};
		\node at (35, 18*\t) [circle, scale=\r, draw=blue,fill=blue] {};
		\node at (15, 19*\t) [circle, scale=\r, draw=black,fill=black] {};
		\node at (30, 19*\t) [circle, scale=\r, draw=blue,fill=blue] {};
		\node at (15, 20*\t) [circle, scale=\r, draw=black,fill=black] {};
		\node at (46, 20*\t) [circle, scale=\r, draw=blue,fill=blue] {};
		\node at (15, 21*\t) [circle, scale=\r, draw=black,fill=black] {};
		\node at (27, 21*\t) [circle, scale=\r, draw=blue,fill=blue] {};
		\node at (15, 22*\t) [circle, scale=\r, draw=black,fill=black] {};
		\node at (20, 22*\t) [circle, scale=\r, draw=blue,fill=blue] {};
		\node at (15, 23*\t) [circle, scale=\r, draw=black,fill=black] {};
		\node at (15, 23*\t) [circle, scale=\r, draw=blue,fill=blue] {};
		\node at (15, 24*\t) [circle, scale=\r, draw=black,fill=black] {};
		\node at (29, 24*\t) [circle, scale=\r, draw=blue,fill=blue] {};
		\node at (15, 25*\t) [circle, scale=\r, draw=black,fill=black] {};
		\node at (22, 25*\t) [circle, scale=\r, draw=blue,fill=blue] {};
		\node at (15, 26*\t) [circle, scale=\r, draw=black,fill=black] {};
		\node at (31, 26*\t) [circle, scale=\r, draw=blue,fill=blue] {};
		\node at (14, 27*\t) [circle, scale=\r, draw=black,fill=black] {};
		\node at (17, 27*\t) [circle, scale=\r, draw=blue,fill=blue] {};
		\node at (14, 28*\t) [circle, scale=\r, draw=black,fill=black] {};
		\node at (30, 28*\t) [circle, scale=\r, draw=blue,fill=blue] {};
		\node at (14, 29*\t) [circle, scale=\r, draw=black,fill=black] {};
		\node at (35, 29*\t) [circle, scale=\r, draw=blue,fill=blue] {};
		\node at (14, 30*\t) [circle, scale=\r, draw=black,fill=black] {};
		\node at (30, 30*\t) [circle, scale=\r, draw=blue,fill=blue] {};
		\node at (14, 31*\t) [circle, scale=\r, draw=black,fill=black] {};
		\node at (17, 31*\t) [circle, scale=\r, draw=blue,fill=blue] {};
		\node at (14, 32*\t) [circle, scale=\r, draw=black,fill=black] {};
		\node at (61, 32*\t) [circle, scale=\r, draw=blue,fill=blue] {};
		\node at (14, 33*\t) [circle, scale=\r, draw=black,fill=black] {};
		\node at (16, 33*\t) [circle, scale=\r, draw=blue,fill=blue] {};
		\node at (14, 34*\t) [circle, scale=\r, draw=black,fill=black] {};
		\node at (12, 34*\t) [circle, scale=\r, draw=blue,fill=blue] {};
		\node at (14, 35*\t) [circle, scale=\r, draw=black,fill=black] {};
		\node at (40, 35*\t) [circle, scale=\r, draw=blue,fill=blue] {};
		\node at (14, 36*\t) [circle, scale=\r, draw=black,fill=black] {};
		\node at (23, 36*\t) [circle, scale=\r, draw=blue,fill=blue] {};
		\node at (13, 37*\t) [circle, scale=\r, draw=black,fill=black] {};
		\node at (39, 37*\t) [circle, scale=\r, draw=blue,fill=blue] {};
		\node at (13, 38*\t) [circle, scale=\r, draw=black,fill=black] {};
		\node at (20, 38*\t) [circle, scale=\r, draw=blue,fill=blue] {};
		\node at (13, 39*\t) [circle, scale=\r, draw=black,fill=black] {};
		\node at (30, 39*\t) [circle, scale=\r, draw=blue,fill=blue] {};
		\node at (13, 40*\t) [circle, scale=\r, draw=black,fill=black] {};
		\node at (18, 40*\t) [circle, scale=\r, draw=blue,fill=blue] {};
		\node at (13, 41*\t) [circle, scale=\r, draw=black,fill=black] {};
		\node at (32, 41*\t) [circle, scale=\r, draw=blue,fill=blue] {};
		\node at (13, 42*\t) [circle, scale=\r, draw=black,fill=black] {};
		\node at (72, 42*\t) [circle, scale=\r, draw=blue,fill=blue] {};
		\node at (13, 43*\t) [circle, scale=\r, draw=black,fill=black] {};
		\node at (29, 43*\t) [circle, scale=\r, draw=blue,fill=blue] {};
		\node at (13, 44*\t) [circle, scale=\r, draw=black,fill=black] {};
		\node at (33, 44*\t) [circle, scale=\r, draw=blue,fill=blue] {};
		\node at (12, 45*\t) [circle, scale=\r, draw=black,fill=black] {};
		\node at (25, 45*\t) [circle, scale=\r, draw=blue,fill=blue] {};
		\node at (12, 46*\t) [circle, scale=\r, draw=black,fill=black] {};
		\node at (13, 46*\t) [circle, scale=\r, draw=blue,fill=blue] {};
		\node at (12, 47*\t) [circle, scale=\r, draw=black,fill=black] {};
		\node at (50, 47*\t) [circle, scale=\r, draw=blue,fill=blue] {};
		\node at (12, 48*\t) [circle, scale=\r, draw=black,fill=black] {};
		\node at (26, 48*\t) [circle, scale=\r, draw=blue,fill=blue] {};
		\node at (12, 49*\t) [circle, scale=\r, draw=black,fill=black] {};
		\node at (27, 49*\t) [circle, scale=\r, draw=blue,fill=blue] {};
		\node at (12, 50*\t) [circle, scale=\r, draw=black,fill=black] {};
		\node at (27, 50*\t) [circle, scale=\r, draw=blue,fill=blue] {};
		\node at (11, 51*\t) [circle, scale=\r, draw=black,fill=black] {};
		\node at (22, 51*\t) [circle, scale=\r, draw=blue,fill=blue] {};
		\node at (11, 52*\t) [circle, scale=\r, draw=black,fill=black] {};
		\node at (16, 52*\t) [circle, scale=\r, draw=blue,fill=blue] {};
		\node at (11, 53*\t) [circle, scale=\r, draw=black,fill=black] {};
		\node at (18, 53*\t) [circle, scale=\r, draw=blue,fill=blue] {};
		\node at (10, 54*\t) [circle, scale=\r, draw=black,fill=black] {};
		\node at (25, 54*\t) [circle, scale=\r, draw=blue,fill=blue] {};
		\node at (10, 55*\t) [circle, scale=\r, draw=black,fill=black] {};
		\node at (4, 55*\t) [circle, scale=\r, draw=blue,fill=blue] {};
		\node at (10, 56*\t) [circle, scale=\r, draw=black,fill=black] {};
		\node at (71, 56*\t) [circle, scale=\r, draw=blue,fill=blue] {};
		\node at (9, 57*\t) [circle, scale=\r, draw=black,fill=black] {};
		\node at (10, 57*\t) [circle, scale=\r, draw=blue,fill=blue] {};
		\node at (9, 58*\t) [circle, scale=\r, draw=black,fill=black] {};
		\node at (10, 58*\t) [circle, scale=\r, draw=blue,fill=blue] {};
		\node at (7, 59*\t) [circle, scale=\r, draw=black,fill=black] {};
		\node at (17, 59*\t) [circle, scale=\r, draw=blue,fill=blue] {};
		\node at (6, 60*\t) [circle, scale=\r, draw=black,fill=black] {};
		\node at (22, 60*\t) [circle, scale=\r, draw=blue,fill=blue] {};
		
		\addplot[red, mark=none] coordinates {(21, -15*\t) (56, -15*\t)};
		\addplot[red, mark=none] coordinates {(21, -14*\t) (31, -14*\t)};
		\addplot[red, mark=none] coordinates {(18, -13*\t) (50, -13*\t)};
		\addplot[red, mark=none] coordinates {(16, -12*\t) (48, -12*\t)};
		\addplot[red, mark=none] coordinates {(16, -11*\t) (31, -11*\t)};
		\addplot[red, mark=none] coordinates {(15, -10*\t) (29, -10*\t)};
		\addplot[red, mark=none] coordinates {(13, -9*\t) (21, -9*\t)};
		\addplot[red, mark=none] coordinates {(12, -8*\t) (29, -8*\t)};
		\node at (21, -15*\t) [circle, scale=\r, draw=black,fill=black] {};
		\node at (56, -15*\t) [circle, scale=\r, draw=red,fill=red] {};
		\node at (21, -14*\t) [circle, scale=\r, draw=black,fill=black] {};
		\node at (31, -14*\t) [circle, scale=\r, draw=red,fill=red] {};
		\node at (18, -13*\t) [circle, scale=\r, draw=black,fill=black] {};
		\node at (50, -13*\t) [circle, scale=\r, draw=red,fill=red] {};
		\node at (16, -12*\t) [circle, scale=\r, draw=black,fill=black] {};
		\node at (48, -12*\t) [circle, scale=\r, draw=red,fill=red] {};
		\node at (16, -11*\t) [circle, scale=\r, draw=black,fill=black] {};
		\node at (31, -11*\t) [circle, scale=\r, draw=red,fill=red] {};
		\node at (15, -10*\t) [circle, scale=\r, draw=black,fill=black] {};
		\node at (29, -10*\t) [circle, scale=\r, draw=red,fill=red] {};
		\node at (13, -9*\t) [circle, scale=\r, draw=black,fill=black] {};
		\node at (21, -9*\t) [circle, scale=\r, draw=red,fill=red] {};
		\node at (12, -8*\t) [circle, scale=\r, draw=black,fill=black] {};
		\node at (29, -8*\t) [circle, scale=\r, draw=red,fill=red] {};
	\end{axis}

\end{tikzpicture}
\end{figure}
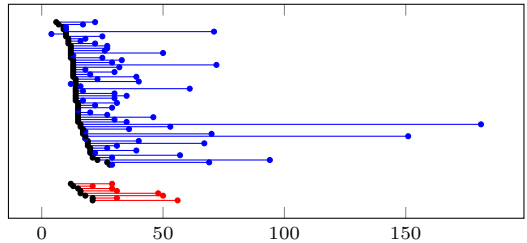

In Figure~\ref{fig:deviances} we consider statistics of the results of runs in which \simpl{} could not derive the exact ground truths as results. Interestingly, the deviance in complexity is in general not larger when \simpl{} completely fails to show an equivalence. Also notably, \simpl's results are in rare cases even simpler than the desired ones.

We omit detailed results for the \textit{QSynth Syntia} and \textit{EA-ED} datasets which contain even more complex expressions (up to $5\,000\,000$ characters per expression); the majority of expressions is solvable, with a few timeouts.



\subsection{Summary}

\simba{} can not only be applied to linear expressions, but also to nonlinear expressions that are reducible to linear ones. We observed that it simplified all expressions in  whose ground truths are linear correctly.

Further, we see that \simpl{} can simplify almost all MBAs to the exact same result as the corresponding ground truth, and it does so quickly (within $50$ ms typically, there are few outlines in the datasets). However, \simpl{} is about ten times slower compared to \simba{} on the datasets. The \textit{QSynth EA} dataset is the hardest to simplify. We want to emphasize that even in case of failure\footnote{Note that ``failure'' means that we cannot verify an MBA's equivalence to the groundtruth using \simpl{} or Z3. However, we checked the solutions numerically for a large number of inputs.} ($\times$), \simpl{}'s results are usually drastically reduced compared to the input MBA and quite close to the corresponding ground truths. We have not observed any expression which could not be simplified significantly.

\simpl{}'s runtimes are, as expected, higher than \simba's, but still very practical. Even for linear MBAs, \simpl{} first has to parse them and figure out that they are linear before they are simplified via a call to \simba. For nonlinear MBAs, it performs additional refinement steps as well as substitutions and calls \simba{} multiple times.

\section{Conclusion}

In this paper we have extended the algorithm presented in~\cite{simba} and its range of possible applications. We have seen in Section~\ref{sec:results} that it is, in the current state of development, already a powerful tool that can simplify a wide variety of expressions.

Our substitution logic and refinement steps make \simpl{} solve MBAs which comparable tools cannot simplify. For instance, many tools have problems with large -- apparently random -- constants, and nonlinear expressions as operands of bitwise operators. Besides, \simpl{} can solve MBAs whose linear parts are obscured via expansion of products to a certain degree.

We experienced the biggest challenge with MBAs that use a high number of variables and cannot be split into expressions with fewer variables in a meaningful way, and with MBAs that have constants within bitwise operations. The substitution logic may be faced with problems when substituting all constants by variables, and furthermore doing so will obscure the relation between the constants. Hence one may hope that the refinement techniques will make the constants vanish.

We believe \simba{} and \simpl{} are valuable tools for the analysis of code obfuscated with MBA expressions. They are straightforward to use and can be easily deployed in program analysis frameworks.

In spite of the challenges mentioned above, we are convinced to have shown that in general every MBA is solvable, independently of its type or complexity. We summarize our main contributions in this paper:

\begin{enumerate}
	\item We have extended the linear simplifier \simba{} such that it finds linear combinations of an unnegated and a negated bitwise expression as well as of two negated ones, closing the gap to be able to find simplest solutions in all cases for two variables.
	
	\item We have introduced various metrics and exhaustively search the space of all possible solutions in order to find the one minimizing a given metric. We have emphasized that result vectors can often be decomposed in a large number of ways and those may imply solutions of very different complexity.
	
	\item We have extended \simba{} to find simple solutions for any number of variables, including more than three.
		
	\item In this course, we have described how to generate bitwise expressions that fit given truth value vectors.
	
	\item We have introduced the powerful, open-source algorithm \simpl{} for simplifying nonlinear MBAs of any kind, applicable to a wide range of inputs MBAs.
	
	\item In contrast to existing algebraic simplifiers, \simpl{} is able to simplify expressions that have constants or arithmetic operations within bitwise operations.
	
	
	\item We have given arguments suggesting that any kind of MBA is simplifiable by algebraic means. \simpl{} can solve all public datasets known to us and has good performance, as our experiments show.


\end{enumerate}

\section*{Data Availability}
\vspace{-1mm}
\simpl's source code is on Github~\cite{gh-gamba}. Datasets used for experiments are referenced in Subsection~\ref{sec:results}.
\vspace{-1mm}

\bibliographystyle{plain}
\bibliography{refs}

\begin{thebibliography}{10}

\bibitem{syntia}
Tim Blazytko, Moritz Contag, Cornelius Aschermann, and Thorsten Holz.
\newblock Syntia. {S}ynthesizing the semantics of obfuscated code.
\newblock In {\em Proc. 26th USENIX Security Symposium}, pages 643--659,
  Vancouver, August 2017.

\bibitem{msynth}
Tim Blazytko and Moritz Schloegel.
\newblock msynth.
\newblock \url{https://github.com/mrphrazer/msynth}, 2020.

\bibitem{tigress}
Christian Collberg.
\newblock Tigress.
\newblock \url{https://tigress.wtf/}.

\bibitem{qsynth}
Robin David, Luigi Coniglio, and Mariano Ceccato.
\newblock {QSynth}. {A} program synthesis based approach for binary code
  deobfuscation.
\newblock In {\em Workshop on Binary Analysis Research (BAR), NDSS Symposium
  2020}, San Diego, February 2020.

\bibitem{eyrolles}
Ninon Eyrolles.
\newblock {\em Obfuscation with Mixed {B}oolean-Arithmetic Expressions.
  {R}econstruction, Analysis and Simplification Tools}.
\newblock PhD thesis, Universit\'{e} Paris-Saclay, France, 2017.

\bibitem{sspam}
Ninon Eyrolles, Louis Goubin, and Marion Videau.
\newblock Defeating {MBA}-based obfuscation.
\newblock In {\em Proc. 2nd ACM Workshop on Software PROtection, SPRO'16},
  pages 27--38, Vienna, October 2016.

\bibitem{eqsat}
Matteo Favaro and Tim Blazytko.
\newblock Improving {MBA} deobfuscation using equality saturation.
\newblock \url{https://secret.club/2022/08/08/eqsat-oracle-synthesis.html},
  2022.

\bibitem{neureduce}
Weijie Feng, Binbin Liu, Dongpeng Xu, Qilong Zheng, and Yun Xu.
\newblock {NeuReduce}. {R}educing mixed {B}oolean-arithmetic expressions by
  recurrent neural network.
\newblock In {\em Findings of the Assoc. for Computational Linguistics}, EMNLP
  2020, pages 635--644, Nov. 2020.

\bibitem{goomba}
Garrett Gu.
\newblock Hands-free binary deobfuscation with {gooMBA}.
\newblock \url{https://hex-rays.com/blog/deobfuscation-with-goomba}, January
  2023.

\bibitem{arybo}
Adrien Guinet, Ninon Eyrolles, and Marion Videau.
\newblock Arybo: {M}anipulation, canonicalization and identification of mixed
  {B}oolean-arithmetic symbolic expressions.
\newblock In {\em GreHack 2016}, Grenoble, France, November 2016.

\bibitem{mba-obfuscator}
Binbin Liu, Weijie Feng, Qilong Zheng, Jing Li, and Dongpeng Xu.
\newblock Software obfuscation with non-linear mixed boolean-arithmetic
  expressions.
\newblock In {\em Proc. Int. Conference on Information and Communications
  Security, ICICS'21}, volume 12918 of {\em LNCS}, pages 276--292, Chongqing,
  China, September 2021. Springer.

\bibitem{mba-blast}
Binbin Liu, Junfu Shen, Jiang Ming, Qilong Zheng, Jing Li, and Dongpeng Xu.
\newblock {MBA-Blast}. {U}nveiling and simplifying mixed {B}oolean-arithmetic
  obfuscation.
\newblock In {\em Proc. 30th USENIX Security Symposium}, pages 1701--1718, Aug.
  2021.

\bibitem{mba-flatten}
Binbin Liu, Qilong Zheng, Jiang Ming, and Dongpeng Xu.
\newblock An in-place simplification on mixed boolean-arithmetic expressions.
\newblock {\em Security and Communication Networks}, 2022:1--14, September
  2022.

\bibitem{mccluskey}
Edward~J. McCluskey.
\newblock Minimization of {B}oolean functions.
\newblock {\em The Bell System Technical Journal}, 35(6):1417--1444, 1956.

\bibitem{xyntia}
Gr\'{e}goire Menguy, S\'{e}bastien Bardin, Richard Bonichon, and Cauim
  de~Souza~Lima.
\newblock {AI}-based blackbox code deobfuscation. {U}nderstand, improve and
  mitigate.
\newblock {\em CoRR}, abs/2102.04805, Feb. 2021.

\bibitem{mougey14}
Camille Mougey and Francis Gabriel.
\newblock {DRM} obfuscation versus auxiliary attacks.
\newblock
  \url{https://recon.cx/2014/slides/recon2014-21-mougey-camille-francis-gabriel-DRM-obfuscation-versus-auxiliary-attacks-slides.pdf},
  June 2014.
\newblock REcon'14, Montreal.

\bibitem{okane11}
Philip O'Kane, Sakir Sezer, and Kieran McLaughlin.
\newblock Obfuscation: The hidden malware.
\newblock {\em IEEE Security \& Privacy}, 9(5):41--47, Sep. 2011.

\bibitem{quine}
W.~V. Quine.
\newblock The problem of simplifying truth functions.
\newblock {\em The American Mathematical Monthly}, 59(8):521--531, 1952.

\bibitem{simba}
Benjamin Reichenwallner and Peter Meerwald-Stadler.
\newblock Efficient deobfuscation of linear mixed boolean-arithmetic
  expressions.
\newblock In {\em Proc. 2022 ACM Workshop on Research on offensive and
  defensive techniques in the context of Man-At-The-End attacks}, CheckMATE'22,
  pages 19--28, Los Angeles, November 2022.

\bibitem{gh-gamba}
Benjamin Reichenwallner and Peter Meerwald-Stadler.
\newblock {GAMBA} code and dataset.
\newblock \url{https://github.com/DenuvoSoftwareSolutions/GAMBA}, 2023.

\bibitem{z3}
Microsoft Research.
\newblock Z3.
\newblock \url{https://github.com/Z3Prover/z3}.

\bibitem{stoke}
Eric Schkufza, Rahul Sharma, and Alex Aiken.
\newblock Stochastic superoptimization.
\newblock In {\em ACM SIGPLAN Notices}, volume~48, pages 305--316, April 2013.

\bibitem{loki}
Moritz Schloegel, Tim Blazytko, Moritz Contag, Cornelius Aschermann, Julius
  Basler, Thorsten Holz, and Ali Abbasi.
\newblock {LOKI}. {H}ardening code obfuscation against automated attacks.
\newblock In {\em 31st USENIX Security Symposium}, pages 3055--3073, Boston,
  Aug. 2022.

\bibitem{schrijver21}
Antoine~De Schrijver.
\newblock Automated localisation of a mixed boolean arithmetic obfuscation
  window in a program binary.
\newblock Master's thesis, Ghent University, Belgium, 2021.

\bibitem{schrittwieser17}
Sebastian Schrittwieser, Stefan Katzenbeisser, Johannes Kinder, Georg
  Merzdovnik, and Edgar Weippl.
\newblock Protecting software through obfuscation: Can it keep pace with
  progress in code analysis?
\newblock {\em ACM Computing Surveys}, 49(1):1--37, March 2017.

\bibitem{mba-solver}
Dongpeng Xu, Binbin Liu, Weijie Feng, Jiang Ming, Qilong Zheng, and Qiaoyan Yu.
\newblock Boosting {SMT} solver performance on mixed-bitwise-arithmetic
  expressions.
\newblock In {\em Proc. 42nd ACM SIGPLAN Int. Conference on Programming
  Language Design and Implementation, PLDI'21}, pages 651--664, Virtual,
  Canada, June 2021.

\bibitem{zhou}
Yongxin Zhou, Alec Main, Yuan~X. Gu, and Harold Johnson.
\newblock Information hiding in software with mixed {B}oolean-arithmetic
  transforms.
\newblock In {\em Proc. 8th Int. Workshop on Information Security Applications,
  WISA'07}, volume 4867 of {\em LNCS}, pages 61--75, Jeju Island, Korea, August
  2007. Springer.

\end{thebibliography}

\end{document}